\pdfoutput=1

\RequirePackage{fix-cm}

\documentclass[twocolumn,epjc3]{svjour3}

\smartqed

\RequirePackage{graphicx}
\RequirePackage{subfigure}
\RequirePackage{mathptmx}
\RequirePackage{flushend}
\RequirePackage{latexsym}
\RequirePackage[numbers,sort&compress]{natbib}
\RequirePackage[colorlinks,citecolor=blue,urlcolor=blue,linkcolor=blue]{hyperref}
\RequirePackage{xspace}
\RequirePackage{tikz}
\RequirePackage{xcolor}
\RequirePackage{multirow}
\RequirePackage{amsmath}
\definecolor{lime}{HTML}{A6CE39}
\DeclareRobustCommand{\orcidicon}{
	\begin{tikzpicture}
	\draw[lime, fill=lime] (0,0)
	circle [radius=0.16]
	node[white] {{\fontfamily{qag}\selectfont \tiny ID}};
	\draw[white, fill=white] (-0.0625,0.095)
	circle [radius=0.007];
	\end{tikzpicture}
	\hspace{-2mm}
}
\foreach \x in {A, ..., Z}{\expandafter\xdef\csname orcid\x\endcsname{\noexpand\href{https://orcid.org/\csname orcidauthor\x\endcsname}
			{\noexpand\orcidicon}}
}

\newcommand*{\smu}{\ensuremath{\tilde{\mu}}\xspace}

\newcommand*{\stau}{\ensuremath{\tilde{\tau}}\xspace}

\newcommand*{\ninoone}{\ensuremath{\mathchoice%
      {\displaystyle\raise.4ex\hbox{\(\displaystyle\tilde\chi^0_1\)}}%
         {\textstyle\raise.4ex\hbox{\(\textstyle\tilde\chi^0_1\)}}%
       {\scriptstyle\raise.3ex\hbox{\(\scriptstyle\tilde\chi^0_1\)}}%
 {\scriptscriptstyle\raise.3ex\hbox{\(\scriptscriptstyle\tilde\chi^0_1\)}}}\xspace}

\journalname{C. P. C.}

\begin{document}

\title{ Search potential for direct slepton pair production at the CEPC \\with $\sqrt{s}$ = 360 GeV
}

\author{\orcidA{}Feng Lyu\thanksref{e1,addr1}
        \orcidB{}{}Jiarong Yuan\thanksref{addr1,addr2}
        \and
        \orcidC{}{}Huajie Cheng\thanksref{addr3}
        \and
        \orcidG{}{}Jianxiong Wang\thanksref{addr1,addr2}
        \and
        \orcidE{}{}Rabia Hameed\thanksref{addr1,addr2}
        \and
        \orcidF{}{}Da Xu \thanksref{addr1,addr2}
        \and
        \orcidD{}Xuai Zhuang\thanksref{e1,addr1,addr2,addr4}
}

\thankstext{e1}{e-mail: luf@ihep.ac.cn, zhuangxa@ihep.ac.cn (corresponding author)}

\institute{Institute of High Energy Physics, Chinese Academy of Sciences, Yuquan Road 19B, Shijingshan District, Beijing 100049, China \label{addr1}
           \and
           University of the Chinese Academy of Sciences, Yuquan Road 19A, Shijingshan District, Beijing 100049, China \label{addr2}
           \and
           Department of Basic Courses, Naval University of Engineering, Jiefang Blvd 717, Qiaokou District, Wuhan 430033, China \label{addr3}
           \and
           Institute of Physics, Henan Academy of Sciences, Zhengzhou 450046, China 
           \label{addr4}
}

\maketitle

\begin{abstract}

The Circular Electron Positron Collider (CEPC) is designed to operate at key center-of-mass energies: 91.2 GeV as a Z factory for precision Z boson studies, $\approx$ 160 GeV at the threshold for W boson pair production, and 240 GeV as a Higgs factory for copious Higgs boson production. It can be upgraded to 360 GeV (CEPC-360GeV) for enabling top quark-antiquark ($t\bar{t}$) pair production. Beyond enabling high-precision measurements of the Standard Model (SM), CEPC-360GeV is uniquely positioned to perform sear- ches for new physics beyond the SM (BSM), serving as a valuable complement to hadron colliders. This paper presents a sensitivity study on the direct pair production of staus and smuons at the CEPC with $\sqrt{s}$ = 360 GeV, conducted via full Monte Carlo (MC) simulation. Under the assumptions of 1.0 ab$^{-1}$ integrated luminosity and a flat 5\% systematic uncertainty, CEPC-360GeV could potentially discover the combined production of left-handed and right-handed staus up to a mass of 170 GeV (if they exist), or up to 169 GeV for pure left-handed staus and 162 GeV for pure right-handed staus. For direct smuon production, the discovery potential reaches up to 178 GeV under the same conditions.

%The center-of-mass energy of Circular Electron Positron Collider (CEPC) could be upgrade to 360 GeV (CEP C@360GeV) after its ten-year running at 240 GeV. Besides SM precision measurements, CEPC@360GeV also has good potential for BSM physics searches, which is a good complementary for hadron colliders. This paper presents the sensitivity study of direct stau and smuon pair production at CEPC $\sqrt{s}$ = 360 GeV by full Monte Carlo (MC) simulation. With 1.0 ab$^{-1}$ integrated luminosity and the assumption of flat 5\% systematic uncertainty, the CEPC@360 GeV has the potential to discover the production of combined left-handed and right-handed stau up to 170 GeV if exists, or up to 169 (162) GeV for the production of pure left-handed (right-handed) stau; the discovery potential of direct smuon reaches up to 178 GeV with the same assumption.

\end{abstract}

\section*{Declarations}
\label{sec:declarations}
\subsection*{Funding}
This study was supported by the State Key Program of National Natural Science of China (Grant No. 2018YFA0404000 and Grant No. 12135013).

\subsection*{Availability of data and material}
The data used in this study won't be deposited, because this study is a simulation study without any experimental data.

\section{Introduction}
\label{sec:intro}
%%%%%%%%%%%%%%%%%%%%% 
Supersymmetry (SUSY)~\cite{Golfand:1971iw,Volkov:1973ix,Wess:1974tw,Wess:1974jb,Ferrara:1974pu,Salam:1974ig,Martin:1997ns} postulates the existence of a supersymmetric partner for each Standard Model (SM) particle, with the spin of the partner differing by one-half unit. In models conserving $R$-parity~\cite{Farrar:1978xj}, SUSY particles are produced in pairs. 
Following a chain of decays, the lightest supersymmetric particle (LSP) remains stable and emerges as a compelling candidate for dark matter~\cite{Goldberg:1983nd,Ellis:1983ew}.
%After a series of decays, the lightest supersymmetric particle (LSP) remains stable and serves as a viable dark matter candidate~\cite{Goldberg:1983nd,Ellis:1983ew}.

In simplified SUSY models involving only electroweak interactions, the SUSY particle spectrum comprises charginos ($\tilde\chi^{\pm}_i$, where $i=1,2$ and ordered by increasing mass), neutralinos ($\tilde\chi^{0}_i$, $i=1,2,3,4$), charged sleptons ($\tilde{l}$), and sneutrinos ($\tilde{\nu}$). Charginos and neutralinos are mass eigenstates formed by linear combinations of the superpartners of Higgs bosons and electroweak gauge bosons. Sleptons, the superpartners of charged leptons are categorized as left-handed ($\tilde{l}_L$) or right-handed ($\tilde{l}_R$) based on the chiralities of their SM counterparts. The mass eigenstates of sleptons, denoted as $\tilde{l}_i$ ($i=1,2$ and ordered by increasing mass), arise from mixing $\tilde{l}_L$ and $\tilde{l}_R$. In this study, we assume mass degeneracy between $\tilde{l}_L$ and $\tilde{l}_R$.

Models featuring light staus can produce dark matter relic densities consistent with cosmological observations~\cite{Vasquez:2011}. Light sleptons also play a role in neutralino co-annihilation processes in the early universe~\cite{Belanger:2004ag,King:2007vh}. In both gauge-mediated ~\cite{Dine:1981gu,AlvarezGaume:1981wy,Nappi:1982hm} and anomaly-mediated~\cite{Randall:1998uk,Giudice:1998xp} supersymmetry breaking models, slepton masses are  expected to be on the order of 100\,GeV.

%Models with light staus can yield dark matter relic densities consistent with cosmological observations~\cite{Vasquez:2011}. Light sleptons also play a role in neutralino co-annihilation in the early universe~\cite{Belanger:2004ag,King:2007vh}, and light smuons may help explain the observed $(g-2)_\mu$ excess~\cite{DPA:2023PRL}. In gauge-mediated~\cite{Dine:1981gu,AlvarezGaume:1981wy,Nappi:1982hm} and anomaly-mediated~\cite{Randall:1998uk,Giudice:1998xp} SUSY breaking models, slepton masses are typically expected to be of order 100\,GeV.

Previous searches for direct slepton pair production have been performed at the Large Electron-Positron Collider (LEP) and the Large Hadron Collider (LHC). At the LEP, stau (smuon) masses below 96 (99)\,GeV were excluded for mass splittings between the slepton and the LSP greater than 7 (4)\,GeV~\cite{LEPslepton,Heister:2001nk,Heister:2003zk,Abdallah:2003xe,Achard:2003ge,Abbiendi:2003ji}. At the LHC, slepton masses up to 700\,GeV were excluded at the 95\% confidence level for a massless LSP~\cite{SUSY-2018-32,CMS-SUS-17-010}. However, sensitivity diminishes for compressed mass spectra: for example, slepton masses up to 251\,GeV are excluded for a mass splitting of approximately 10\,GeV~\cite{SUSY-2018-16}. Using 139\,fb$^{-1}$ of ATLAS data, stau masses up to 500\,GeV are excluded at the 95\% confidence level~\cite{SUSY-2018-04}; with 77.2\,fb$^{-1}$ of CMS data, masses between 90 and 120\,GeV are excluded under the assumption of a nearly massless LSP~\cite{CMS-SUS-18-006}. For small mass splittings between the slepton and the LSP, sensitivity remains highly limited at hadron colliders, applying to the LHC~\cite{PhysRep1116(2025)261} and the projected performance of the High Luminosity LHC (HL-LHC)~\cite{CERN-2019-007}, which is not expected to resolve this limitation. Challenges posed by such compressed regions at the LHC and HL-LHC arise from difficulties in reconstructing low-energy leptons and severe background contamination, which are limitations that highlight the unique advantages of lepton colliders.

The Circular Electron Positron Collider (CEPC)~\cite{Ruan:2018yrh} is designed to operate at key center-of-mass energies: 91.2 GeV as a Z factory for precision Z boson studies, $\approx$ 160 GeV at the threshold for W boson pair production, and 240 GeV as a Higgs factory for copious Higgs boson production. This can be upgraded to 360 GeV (CEPC-360GeV), which is an energy near the $t\bar{t}$ threshold that enhances sensitivity to physics beyond the SM (BSM) physics. A slepton search using CEPC-240GeV with 5.05\,ab$^{-1}$ of integrated luminosity has already been performed: under this configuration, CEPC-240GeV shows the potential to discover a combined left- and right-handed stau production up to 116\,GeV, or up to 113\,GeV for pure $\tilde{l}_L$ or $\tilde{l}_R$ scenarios. For smuon production, the discovery potential reaches 117\,GeV under the same conditions~\cite{yuanjr}. This paper presents an analysis of the potential for slepton pair production at CEPC-360GeV with 1.0\,ab$^{-1}$ of integrated luminosity.

Similar to the CEPC, the International Linear Collider (ILC) and Future Circular Collider for electrons and positrons (FCC-ee) are proposed electron-positron colliders~\cite{Behnke:2013lya,Gomez-Ceballos:2013zzn}. The CEPC and FCC-ee are circular colliders designed to operate at multiple center-of-mass energies ranging from 90 to 360\,GeV, while the ILC is a linear collider with planned energies spanning 250\,GeV to 1\,TeV. Drawing on prior LEP-based studies~\cite{smiljanic:2020sys}, a conservative systematic uncertainty of 5\% is adopted in this analysis. The results presented in this paper are expected to be largely independent of specific detector, trigger, and data acquisition details, which makes them applicable as references for the FCC-ee and ILC with appropriate luminosity scaling.

\begin{figure}[!htb]
\centering
  \subfigure [direct stau production] {\includegraphics[width=.23\textwidth]{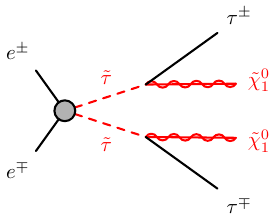}}
  \subfigure [direct smuon production] {\includegraphics[width=.23\textwidth]{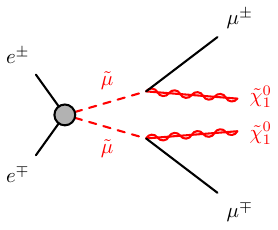}}
\caption{Representative diagram illustrating the pair production of charged staus (smuons) and their subsequent decay into a di-tau (di-muon) final state.}
  \label{fig:feynmanslepton}
\end{figure}

This paper presents sensitivity studies on the direct production of staus and smuons, wherein each slepton is assumed to decay into a lepton ($\tau$ or $\mu$) and the lightest neutralino ($\tilde\chi^0_1$), as illustrated in Figure~\ref{fig:feynmanslepton}. In this scenario, the lightest neutralino acts as the LSP and is assumed to be purely Bino in composition. 

%%%%%%%%%

%\begin{figure}[!htb]
%\centering
%\includegraphics[width=.3\textwidth]{./fig_11.png}
%\caption{ATLAS exclusion limits on slepton (selectron, smuon and stau) masses, assuming %degeneracy of $\slepton_{L}$ and $\slepton_{R}$, and 100\% branching fraction for %$\slepton\to\ell\ninoone$. }
%  \label{fig:atla}
%\end{figure}

\section{Detector, Software and Samples}
\label{sec:samples}

\begin{table}
    \centering
    \caption{Cross-sections of the SM background processes considered at CEPC-360 GeV.}
    \label{tab:xsec}
%    \begin{tabular}{\textwidth}{@{\extracolsep{\fill}}cccc@{}}
\begin{tabular}{p{140 pt}p{70 pt}}
\hline
Process                              & Cross section (fb)\\
\hline
$ZZ$ or $WW ~(\to\mu\mu\nu\nu)$                &    150.87 \\
$\mu\mu$                               &    2430.34 \\
$\nu Z ~(Z\to\mu\mu)$                     &    45.63\\
$ZZ~(\to\mu\mu\nu\nu)$                    &    10.42\\
$WW~(\to\ell\ell\nu\nu)$                  &    281.05\\
$\tau\tau$                             &    2112.22\\
$ZZ$ or $WW~(\to\tau\tau\nu\nu)$              &    145.92\\
$ZZ~(\to\tau\tau\nu\nu)$                  &    10.25 \\
$\nu Z~(Z\to\tau\tau)$                   &    16.15 \\
$\nu\nu H~(H\to$ anything)             &    53.62\\
$t\bar{t}$                                  &    610.93\\
$e\nu W~(W\to\mu\nu)$                    &    365.99\\
$e\nu W~(W\to\tau\nu)$                   &    365.52\\
$eeZ~(Z\to\nu\nu)$                       &    35.66 \\
$eeZ~(Z\to\nu\nu)$ or $e\nu W~(W\to e\nu)$ &  248.41\\
two-photon & 3.96E+8\\
\hline 
\end{tabular}
\end{table}  

The CEPC Conceptual Design Report (CDR) provides a comprehensive introduction to the detector and associated software~\cite{CEPCStudyGroup:2018ghi}. For this Monte Carlo (MC) study, the CEPC baseline detector is employed in full MC simulations, which adhere to the particle flow principle and incorporate an ultra-high-granularity calorimetry system, a low-material silicon tracker, and a 3 Tesla magnetic field. The CEPC baseline tracker comprises a silicon tracking system and  barrel time projection chamber (TPC), which results in nearly 100\% reconstruction efficiencies for muon tracks and tau-decay tracks. In Addition, the momentum resolution of muon tracks reaches the per-mille level for momenta ranging from 10 to 100 GeV in the barrel region.

%The CEPC Conceptual Design Report (CDR) presents the comprehensive introduction of detector and software~\cite{CEPCStudyGroup:2018ghi}. In this MC study, the baseline of CEPC detector is used in the MC full simulation which follows particle flow principle, and uses an ultra high granularity calorimetry system, a low material silicon tracker and a 3 Tesla magnitude field. The CEPC baseline tracker consists of a silicon tracking system and a barrel TPC, which make reconstruction efficiencies of muon and tau-decay tracks nearly 100\%, and the muon track momentum resolution reaches per mille level for the momentum range of 10–100 GeV in the barrel region.

\begin{table*}
  \centering
  \caption{Summary of selection requirements for direct stau production signal regions. $\Delta M$ denotes the mass difference between the $\stau$ and LSP.}
  \label{tab:SRdt}
%\begin{tabular*}{\columnwidth}{@{\extracolsep{\fill}}ccc@{}}
\begin{tabular*}{\textwidth}{@{\extracolsep{\fill}}cccc@{}}
\hline
 SR-$\Delta M^h$ & SR-$\Delta M^m$ & SR-$\Delta M^l$ \\
\hline
%\multicolumn{3}{c}{Preselection:  Events have two OS taus whose $E_\tau$ \textgreater 0.5 GeV}\\
$E_{\tau}$ \textless 40 GeV&\multicolumn{2}{c}{$E_{\tau}$ \textless 15 GeV}\\
$P_{T}$ \textgreater 50 GeV & $P_{T}$ \textgreater 20 GeV & --  \\
2.55 \textless $|\Delta \phi(\tau,recoil)|$ \textless 3.1 & $|\Delta \phi(\tau,recoil)|$ \textless 3.1 & 2.3\textless$|\Delta \phi(\tau,recoil)|$ \textless 2.6\\
- & 0.45\textless $\Delta R(\tau,\tau)$ \textless 1.7 & $\Delta R(\tau,\tau)$ \textgreater 0.45\\
$\Delta R(\tau,recoil)$ \textless 3.2 & $\Delta R(\tau,recoil)$ \textless 3.15 & $\Delta R(\tau,recoil)$ \textless 2.8\\
$M_{\tau\tau}$ \textless 40 GeV &$M_{\tau\tau}$ \textless 25 GeV & 5 GeV\textless$M_{\tau\tau}$ \textless 16 GeV\\
$M_{recoil}$ \textgreater 180 GeV & $M_{recoil}$ \textgreater 280 GeV & $M_{recoil}$ \textgreater 325 GeV\\
\hline
\end{tabular*}
\end{table*}

The software employed in the simulation process is as follows:
SM background samples were generated using Whi- zard 1.95~\cite{Kilian:2011sepTEPJC},   with the version consistent with that used in the official CEPC MC production. 
SUSY signal samples were generated using MadGraph 2.7.3~\cite{Alwall:2014julJHEP} and Pythia8~\cite{Sjostrand:2014zea}.
Particle interactions with detector materials were simulated via MokkaC~\cite{MorasMokka}.
Track reconstruction was performed using MarlinTraking~\cite{Gaede_2014}.
The particle flow algorithm, Arbor~\cite{Ruan:2018yrh}, was utilized to reconstruct all final-state particles.
Lepton identification was implemented using LICH~\cite{Yu:2017mpx}, which is based on multivariate data analysis via TMVA~\cite{hoecker2007tmva}.

In this study, sleptons are pair-produced in electron-posit- ron collisions, with each slepton decaying into a lepton and the lightest neutralino ($\tilde\chi^0_1$) with a 100\% branching ratio. For simplifying the analysis, other supersymmetric particles are excluded from the production or decay chains. The mixing matrices for the scalar tau and muon sectors are considered antidiagonal, precluding mixed production modes. Signal samples for direct stau (smuon) production are parametrized as functions of the masses of $\tilde{\tau}$ ($\tilde{\mu}$) and LSP. The $\tilde{\tau}$ ($\tilde{\mu}$) mass range is constrained between the LEP limits~\cite{LEPslepton,Heister:2001nk,Heister:2003zk,Abdallah:2003xe,Achard:2003ge,Abbiendi:2003ji} and CEPC beam energy, spanning 80--179\,GeV. Here, left- and right-handed slepton superpartners are assumed to be mass-degenerate. Reference points with $\tilde{\tau}$ ($\tilde{\mu}$) masses of 160 (170)\,GeV and $\tilde\chi^0_1$ masses of 50, 110 and 150 (30, 120 and 160)\,GeV are used to illustrate key analysis features. Theoretical cross sections for slepton pair production at CEPC- 360GeV are calculated at leading order (LO) using \texttt{MadGraph} ~\cite{Alwall:2014julJHEP}, yielding 42.22\,fb for $\tilde{\tau}$ (160\,GeV) and 22.34\,fb for $\tilde{\mu}$ (170\,GeV).

In this analysis, SM background processes include both fully decaying $t\bar{t}$ events and non-$t\bar{t}$ processes featuring two leptons (taus, muons, or electrons) with large recoil mass. The background categories comprise $t\bar{t}$ production, two-fermion processes, four-fermion processes, and selected Higgs processes. The Higgs background is restricted to the $\nu\nu H$ channel, encompassing all possible Higgs decay modes. The two-fermion processes include the $\mu\mu$ and $\tau\tau$ final states, while the four-fermion processes include $ZZ$, $WW$, single-$Z$, single-$W$, and mixed $Z$/$W$ production. 
The two-photon background arises from processes including ee $\to$ eeee, ee $\to$ $\mu\mu$ee, and ee $\to$ $\tau\tau$ee.
The labels for four-fermion processes in Table~\ref{tab:xsec} are defined solely by the distinct manner in which the four final-state fermions associate with the underlying Z and W boson combinations. All such processes incorporate the full set of contributions, including the s-channel, t-channel, and their interference, as required by the SM. In addition, the processes listed in the table are mutually exclusive, with no overlapping event contributions. 
The two-photon background were generated using the Feynman diagram calculation (FDC) ~\cite{Wang:2004du}, where the two t-channel photon exchange diagrams provide the leading-order contribution for the beamstrahlung effect. Therefore, higher-order contributions from beamstrahlung are omitted here, and a more precise investigation can be performed if required in the future.
All simulated samples are normalized to an integrated luminosity of 1.0\,ab$^{-1}$. The cross sections of the considered background processes are summarized in Table~\ref{tab:xsec}. 

%We used CEPC official SM backgroud production which their Feynman diagrams corresponding to various SM background classifications are listed one by one in the appendix of the CEPC note ~\cite{cepc:note}, and there is no overlap.

\section{Search for direct slepton production}
\label{sec:searchdl}
Event reconstruction includes three main components: track reconstruction, particle flow reconstruction, and compound physics object reconstruction. Tracks are reconstructed from detector hits using \texttt{Clupatra}~\cite{Gaede_2014}. Particle flow reconstruction combines track and calorimeter information to reconstruct individual particle-level objects, which are then used to build compound physics objects such as converted photons, taus, and jets. The identification efficiency for light leptons (electrons and muons) exceeds 99.5\% for particles with energies above 2\,GeV~\cite{Yu:2017mpx}. For electrons and muons with energies below 1\,GeV, particularly those in the barrel edge or barrel-endcap transition regions, the identification efficiency is approximately 90\%.

%The recoil system consists of all the particles except the two opposite sign (OS) charged leptons, which including invisible particles such as neutrinos and neutralinos. Without considering the beam energy spread, the resolution of the reconstructed recoil mass is between 300 and 400 MeV~\cite{Kilian:2011sepTEPJC}.

The CEPC exhibits distinct properties, including precise energy-momentum conservation (arising from its clean, symmetric initial state), reduced background noise, and a well-defined center-of-mass frame. Variables associated with invariant mass, lepton/photon momentum and energy, angular distributions, and lepton flavor or charge are effective in this context: they capitalize on the clean initial state and predictable final states of the collider to probe new physics with unparalleled precision. Building on this detailed characterization of lepton collider kinematics, the following variables efficiently discriminate signal events from SM backgrounds:

\begin{itemize}
\item $|\Delta \phi(\ell,recoil)|$, the difference of azimuth between one lepton and the recoil system.
\item $|\Delta \phi(\ell,\ell)|$, the difference of azimuth between two leptons.
\item $\Delta R(\ell,recoil)$, the angular distance between one lepton and the recoil system. \footnote{
	$\Delta R = \sqrt{(\Delta\eta)^2+(\Delta\phi)^2}$, where $\eta$ is the pseudorapidity which defined in terms of the polar angle $\theta$ by $\eta=-\ln\tan(\theta/2)$ and $\phi$ is the azimuthal angle.
	}
\item $\Delta R(\ell,\ell)$, the angular distance between two leptons.
\item $E_{\ell}$, the energy of one lepton.
\item $P_T$, the sum of the transverse momentum of two leptons.
\item $M_{\ell\ell}$, the invariant mass of two leptons.
\item $M_{recoil}$, the recoil mass of the di-lepton pair, as defined in Eq (\ref{eq:recoil:recoilmass}), where $\l$ represents a muon (tau) track originating from smuon (stau) pair production.
\begin{equation}
M_{\text{recoil}} =\sqrt{(\sqrt{s} - E_{\l_{1}}-E_{\l_{2}} )^2 - | \vec{p}_{\l_{1}} + \vec{p}_{\l_{2}} |^2}
%M_{\text{recoil}} =\sqrt{(\sqrt{s} - E_{\l^{-}}-E_{\l^{+}} )^2 - | \vec{p}_{\l^{-}} + \vec{p}_{\l^{+}} |^2}
\label{eq:recoil:recoilmass}
\end{equation}

\end{itemize}

\begin{figure*}[!htp]
  \centering
    \subfigure [ $E_{\tau1}$] {\includegraphics[width=.48\textwidth]{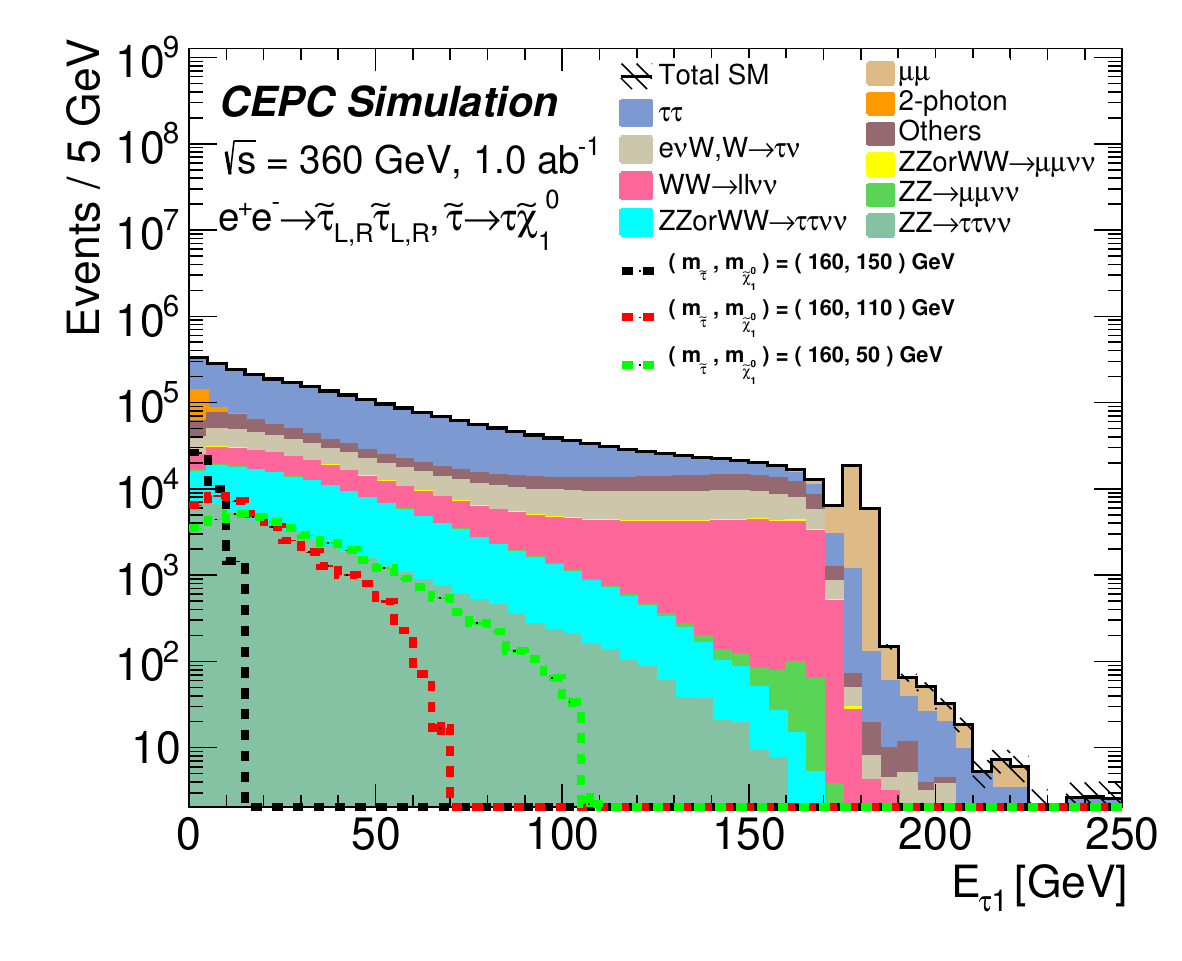}}
    \subfigure [ $P_{T}$]{\includegraphics[width=.48\textwidth]{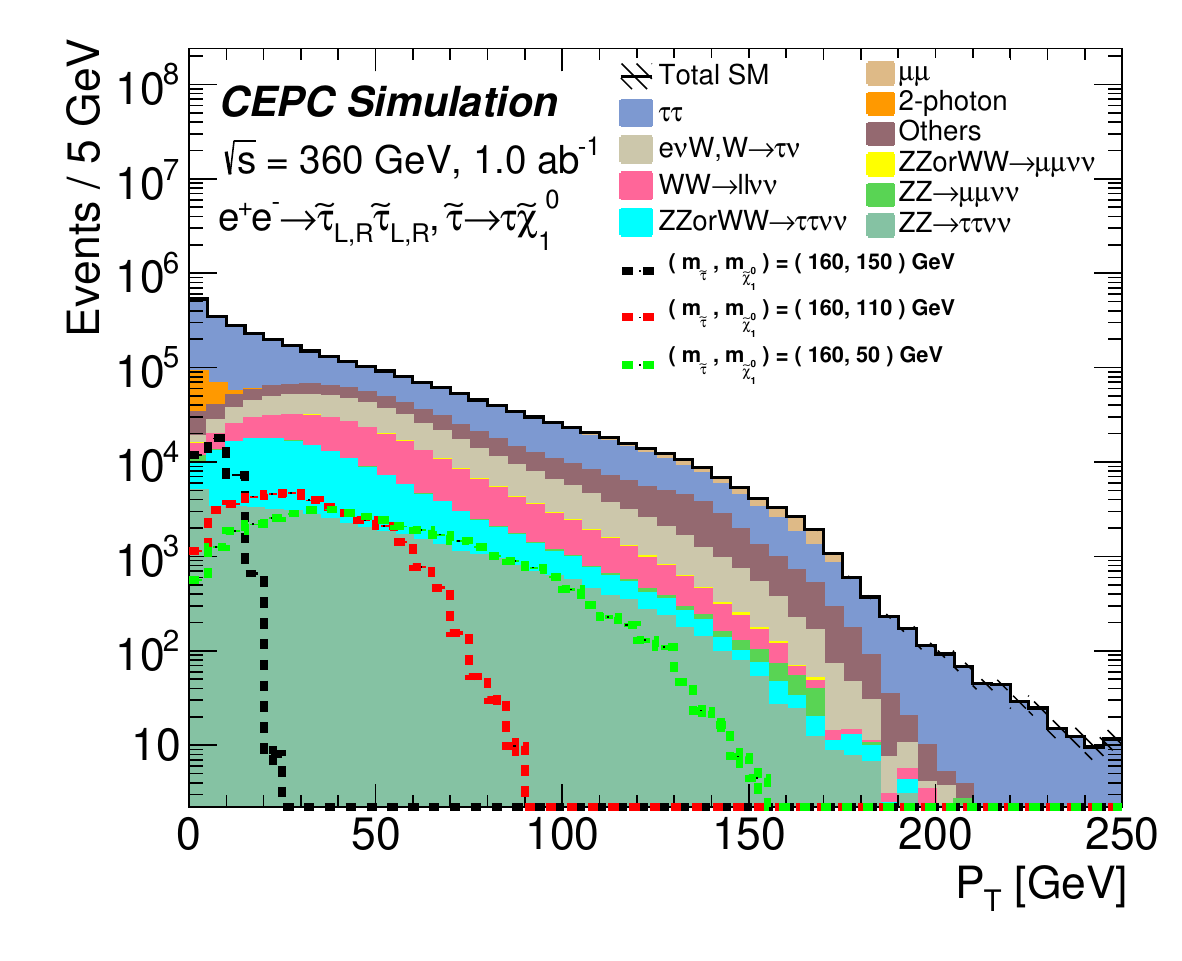}}
     \subfigure [ $M_{\tau\tau}$] {\includegraphics[width=.48\textwidth]{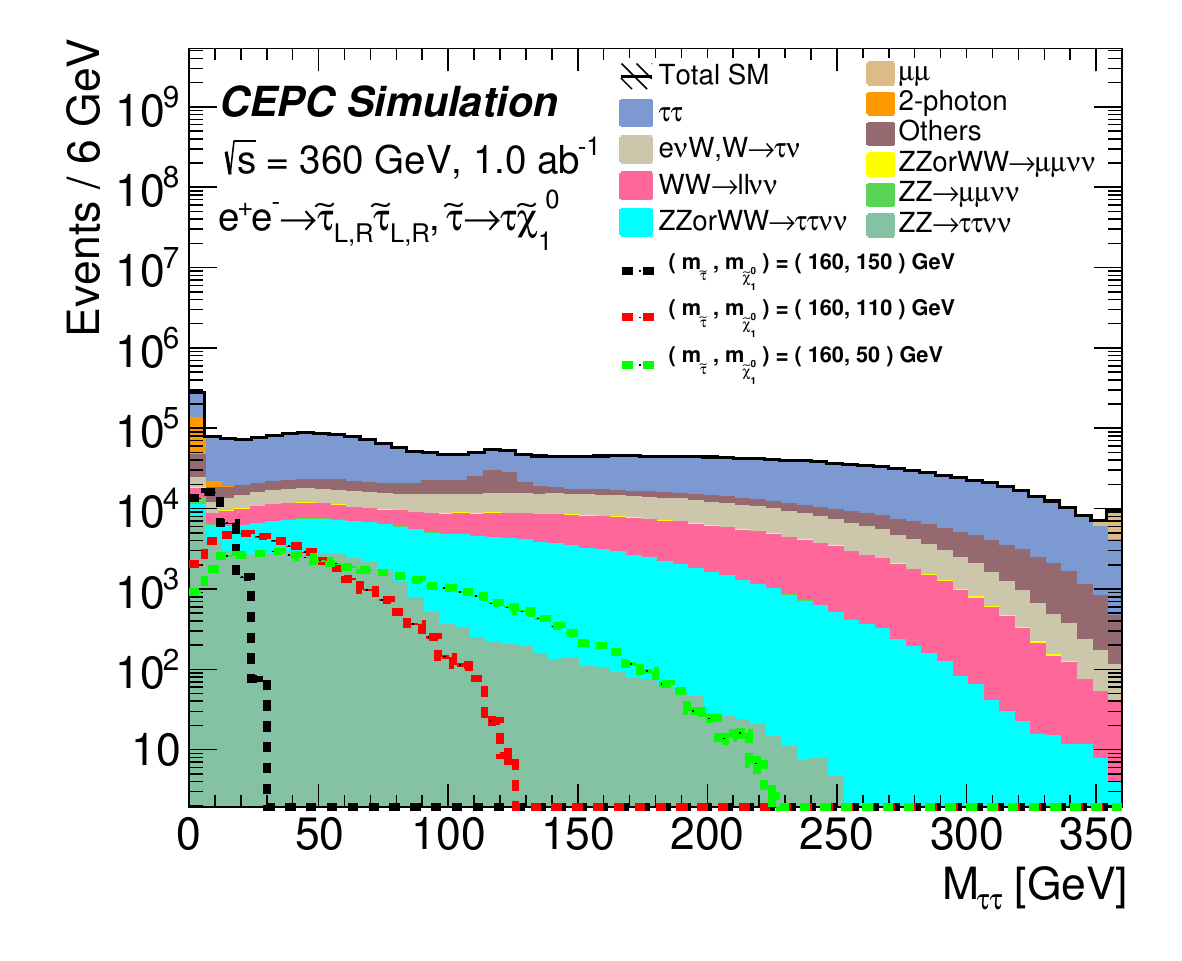}}
    \subfigure [ $M_{recoil}$] {\includegraphics[width=.48\textwidth]{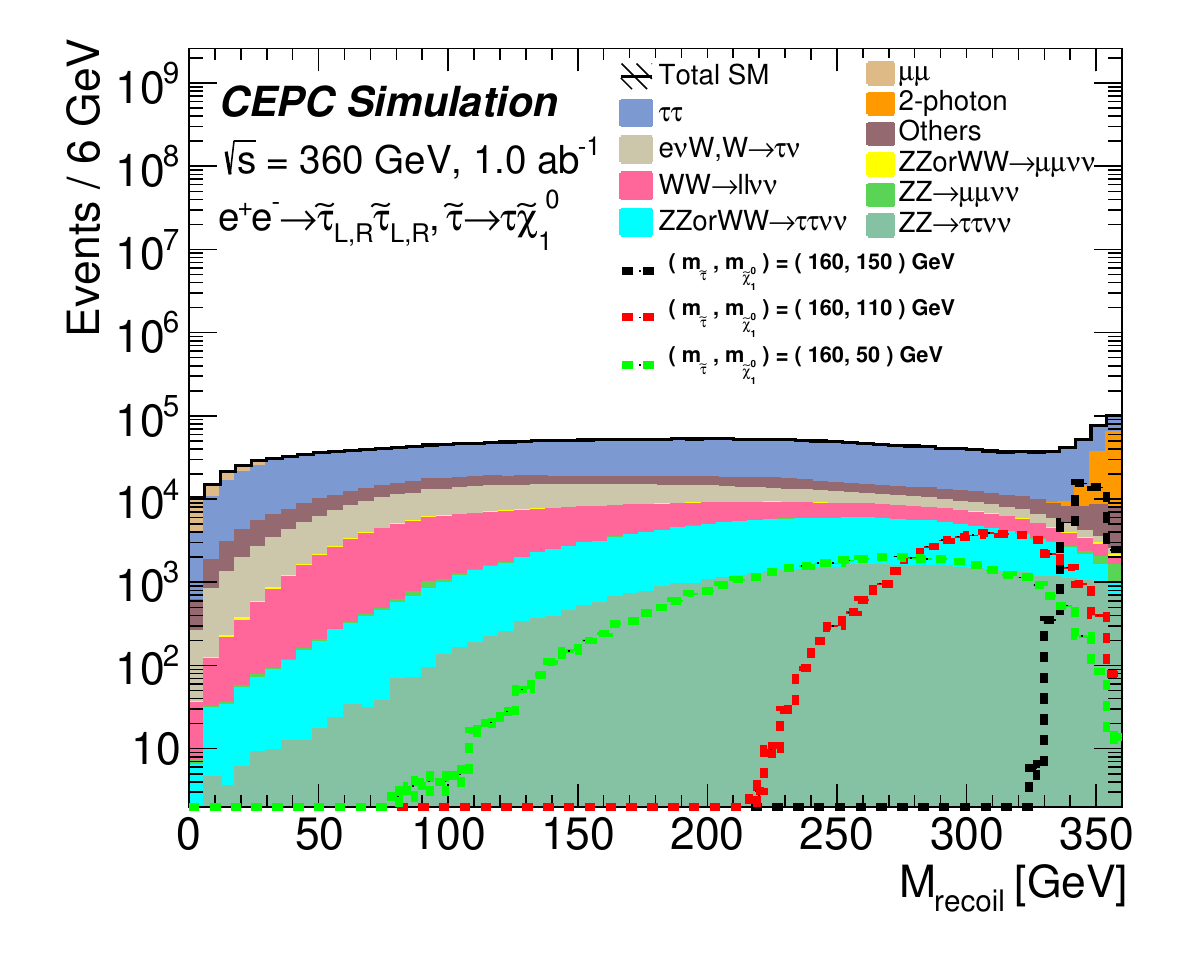}}
    \caption{The kinematic distributions for direct stau pair production after the two OS tau selection with $E_{\tau}$ larger than 2.5 GeV. The stacked histograms show the expected SM background. The distributions of three SUSY reference points are shown as dashed lines.}
    \label{fig:nm0dt}
  \end{figure*}

\begin{figure*}[!htp]
  \centering
    \subfigure [$M_{recoil}$] {\includegraphics[width=.48\textwidth]{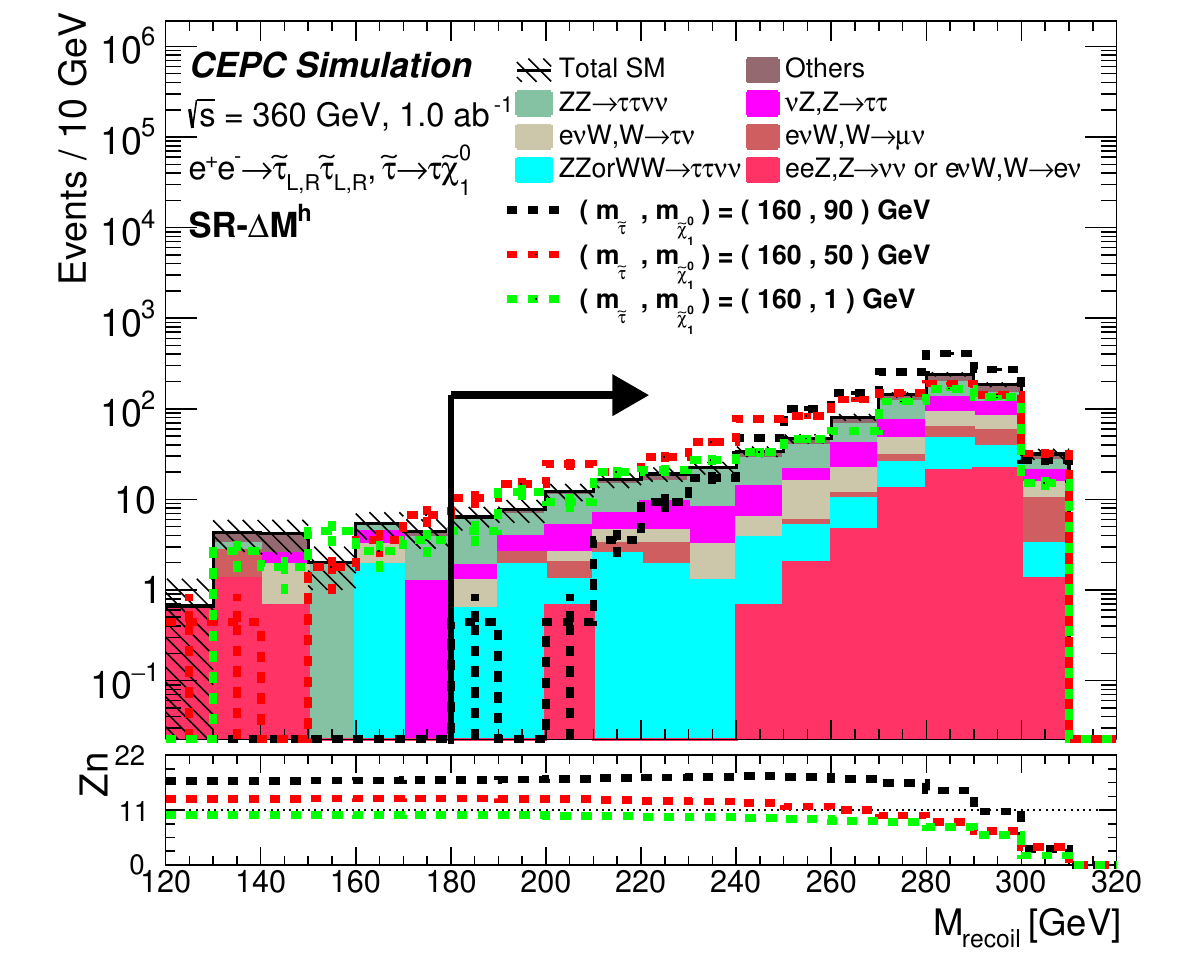}}
    \subfigure [$M_{\tau\tau}$]{\includegraphics[width=.48\textwidth]{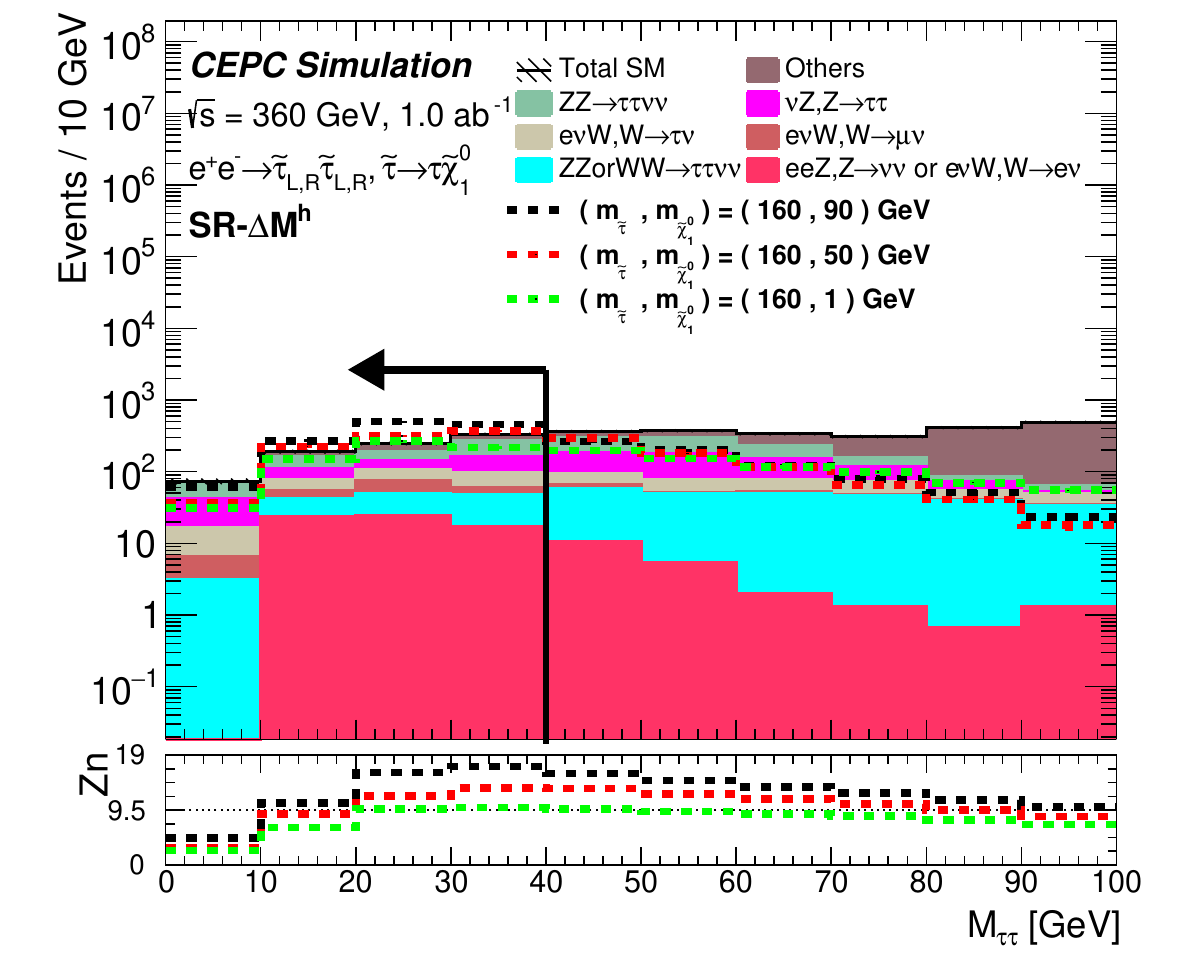}}
    \subfigure [$M_{recoil}$] {\includegraphics[width=.48\textwidth]{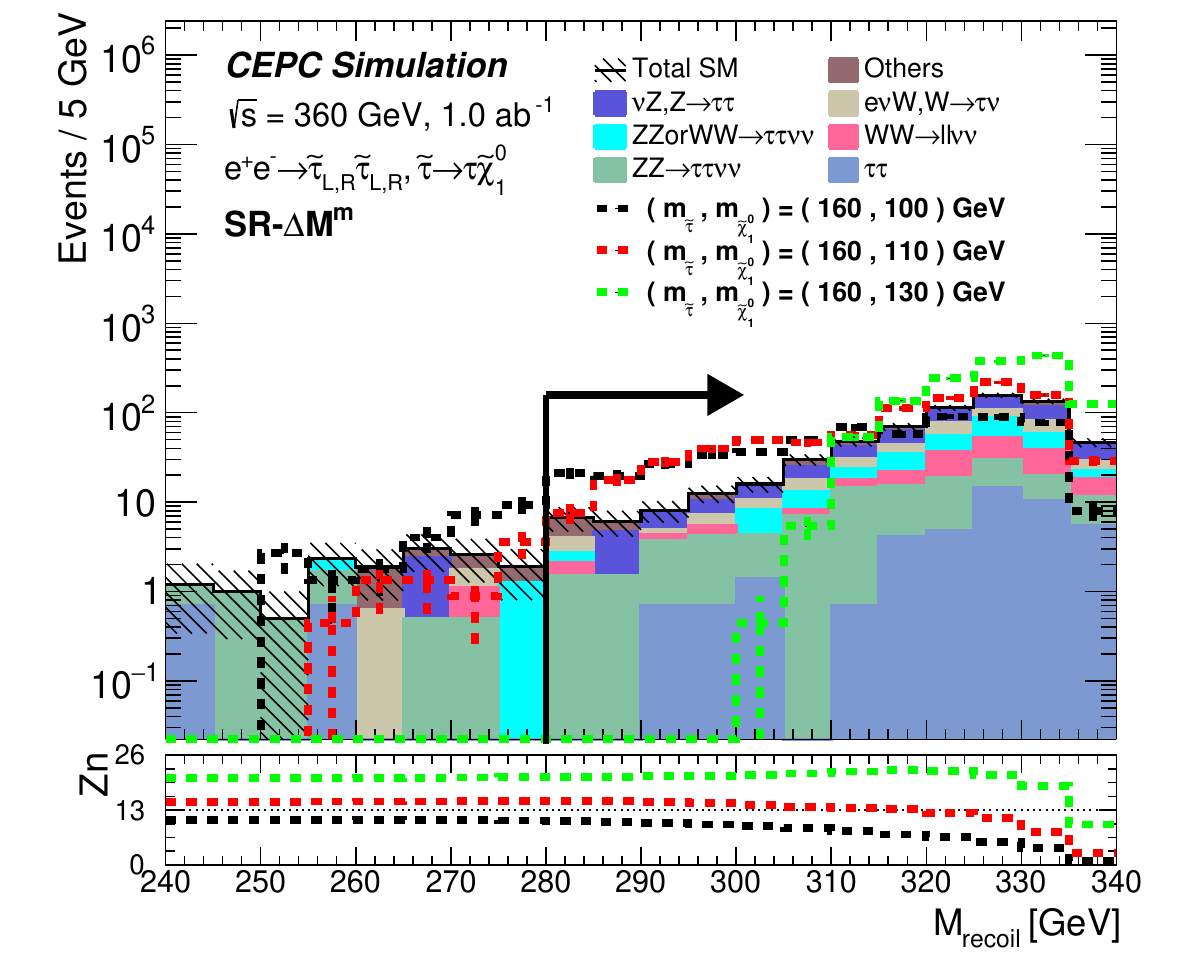}}
    \subfigure [$M_{\tau\tau}$] {\includegraphics[width=.48\textwidth]{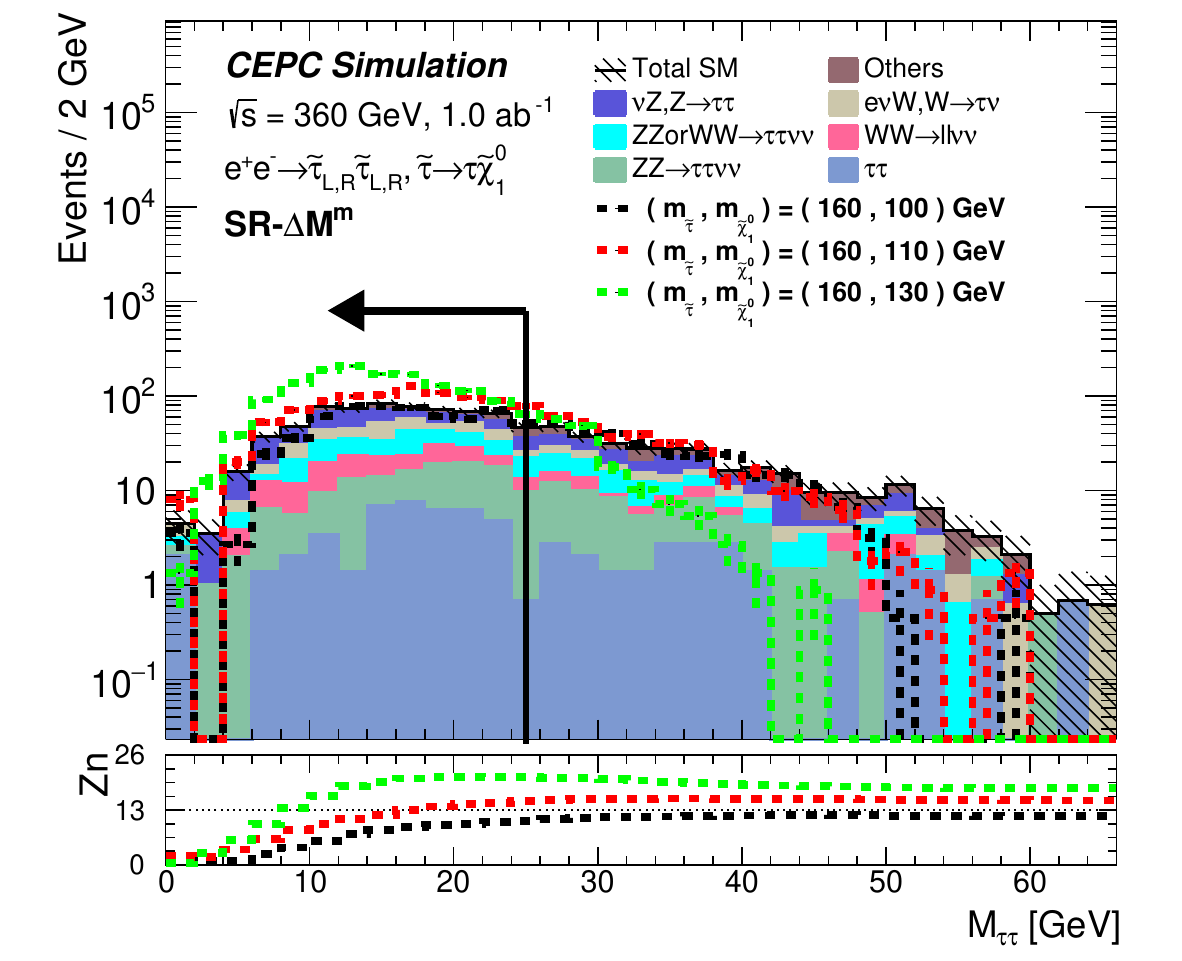}}
    \subfigure [$M_{recoil}$] {\includegraphics[width=.48\textwidth]{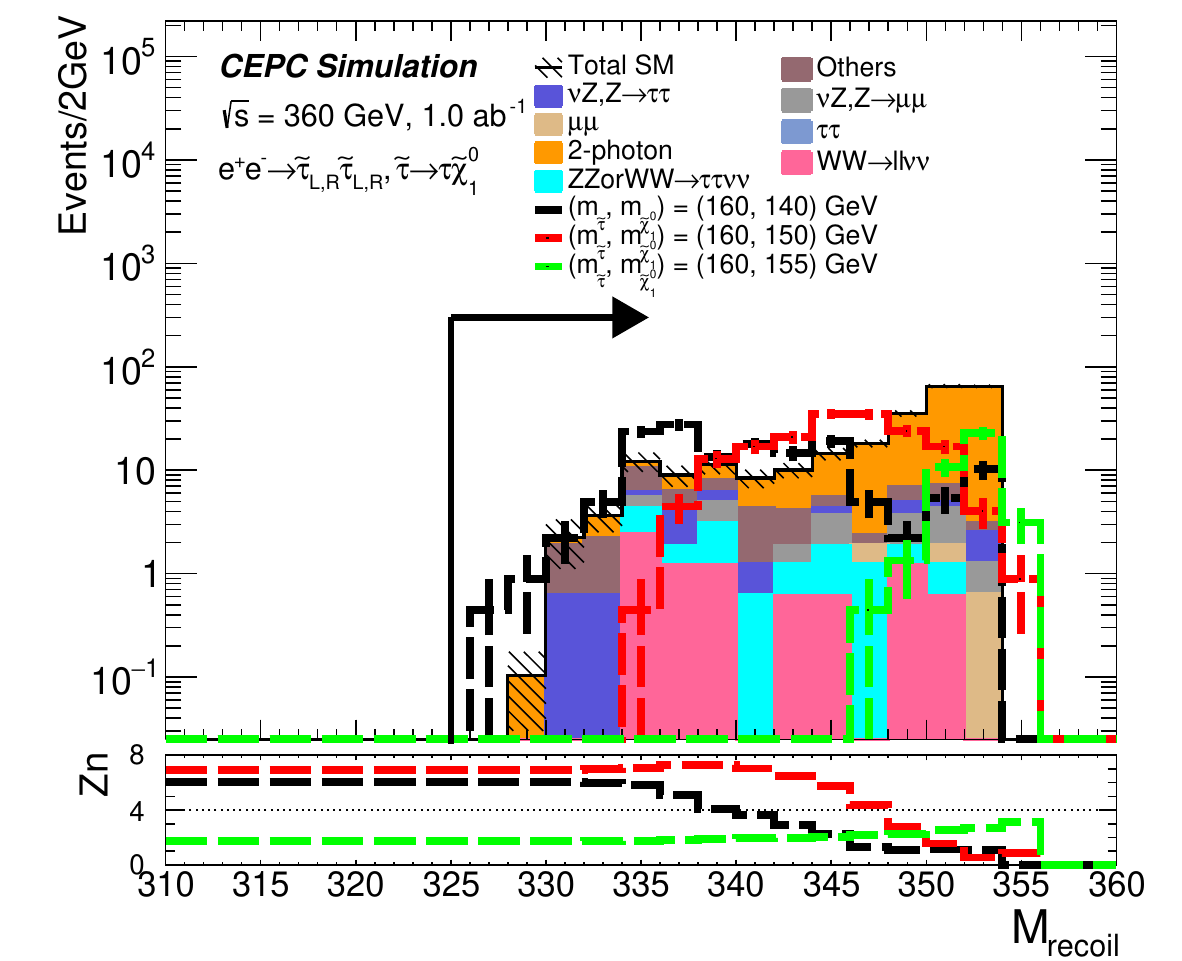}}
    \subfigure [$M_{\tau\tau}$] {\includegraphics[width=.48\textwidth]{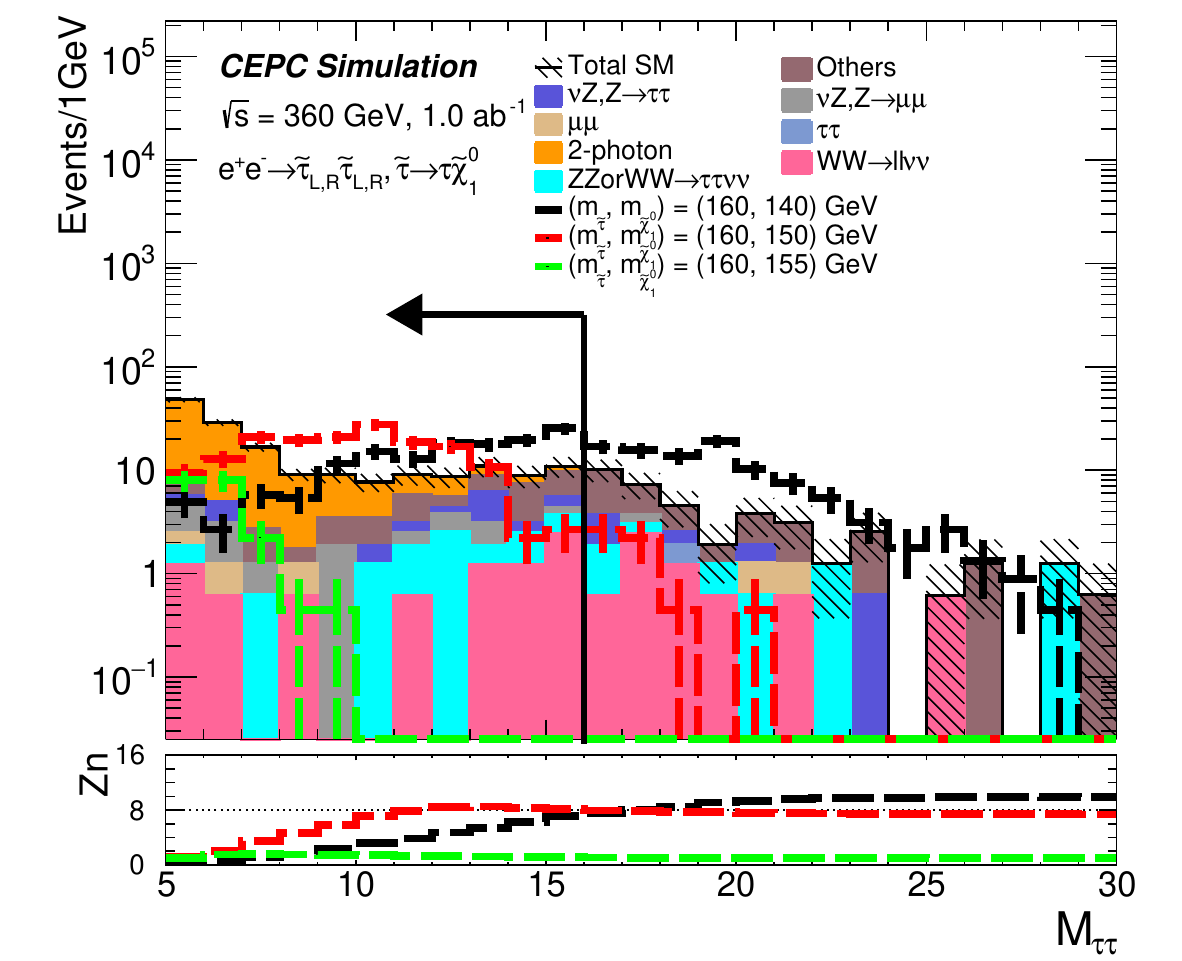}}
    \caption{"N-1" distributions after signal region requirements for direct stau pair production. All signal region requirements are applied except for the variable shown. The stacked histograms show the expected SM backgrounds. The distributions from SUSY reference points are shown as dashed lines. The lower pad is the sensitivity Zn calculated with a statistical uncertainty and a 5\% flat systematic uncertainty.}
    \label{fig:nm1dt}
  \end{figure*}

To mitigate the effect of potential correlations between variables, the optimization of signal regions relies on an N-dimensional scan over the aforementioned kinematic variables, achieved by applying varying cut values. The cut combination yielding the highest signal sensitivity is selected and further validated by reassessing kinematic distributions within the relevant regions. Ultimately, signal regions are defined using the final selected kinematic criteria. To estimate the sensitivity to the signal, the median significance is employed, because it offers a more accurate approximation of the true significance in regions where the number of signal events is non-negligible. This significance, denoted as $Z_\mathrm{n}$, is calculated using the method described in Ref.~\cite{Cowan:2012ds}, as shown in Equation~(\ref{equ:Zn}).

\begin{small}
\begin{equation}
  Zn=\left[2\left((s+b)\ln\left[\frac{(s+b)(b+\sigma_b^2)}{b^2+(s+b)\sigma_b^2}\right]-\frac{b^2}{\sigma_b^2}\ln\left[1+\frac{\sigma_b^2s}{b(b+\sigma_b^2)}\right]\right)\right]^{1/2} \label{equ:Zn}
\end{equation}
\end{small}

In the calculation of Zn, both statistical uncertainty and 5\% flat systematic uncertainty are incorporated. This approach is widely adopted in optimization studies for new physics searches across experiments such as ATLAS.

\subsection{Search for direct stau production}
\label{subsec:searchdt}
For preselection in the search for direct stau production, the following criteria are applied: the two most energetic tracks are required for exactly opposite charges, with each satisfying an energy threshold exceeding 2.5 GeV and an absolute pseudorapidity less than 2.5; this criterion is implemented to suppress the two-photon background. In addition, at least one of these two tracks must originate from a hadronic tau decay.
The negatively charged track is identified as a tau lepton, while the positively charged track is identified as an anti-tau lepton.
Key kinematic distributions after preselection are presented in Figure~\ref{fig:nm0dt}, demonstrating strong discrimination power between signal events and SM background processes.

Signal behaviors depend on the mass splitting between \stau and \ninoone ($\Delta M$), as shown in Figure \ref{fig:nm0dt}, therefore, three signal regions (SRs) are designed to cover the whole \stau-\ninoone mass parameter space. SR-$\Delta M^h$ encompasses the high $\Delta M$ region, SR-$\Delta M^m$ covers the medium $\Delta M$ region, and SR-$\Delta M^l$ spans the low $\Delta M$ region.
The definitions of these signal regions are summarized in Table \ref{tab:SRdt}.
An upper cut on the tau energy ($E_{\tau}$) is applied to suppress $\mu\mu$, Z or W mixing, WW, e$\nu$W and eeZ processes.
A lower cut on the sum of the transverse momentum of two leptons ($P_{T}$) is implemented to reduce backgrounds from $\tau\tau$, $\mu\mu$ and $\nu$Z processes. Given the signal topology, most signal events exhibit a large recoil mass; thus, a lower cut on the invariant mass of the recoil system ($M_{recoil}$) is used to reject $\tau\tau$ and e$\nu$W processes, as well as other SM processes lacking significant recoil mass.
Upper cuts on the $\tau\tau$ invariant mass ($M_{\tau\tau}$) further suppress $\tau\tau$ and e$\nu$W backgrounds, whereas a lower cut on $M_{\tau\tau}$ in the SR-$\Delta M^l$ region is implemented for reducing the two-photon background. Additional suppression of $\tau\tau$ and $\mu\mu$ processes is achieved via selections on $|\Delta \phi(\tau,\tau)|$ and $|\Delta \phi(\tau,recoil)|$.
Selections on $\Delta R(\tau,\tau)$ are employed to further reject $\tau\tau$, ZZ and single-Z processes.
The same signal regions are used for both left-handed and right-handed staus, as their kinematic behaviors are nearly identical, with only small differences in production cross sections.

The kinematic distributions of $M_{recoil}$ and $M_{\tau\tau}$ after applying signal region requirements, except those on the variable to be shown are presented in Figure \ref{fig:nm1dt}.
The lower panel of each plot displays the expected sensitivity Zn as a function of $M_{recoil}$ and $M_{\tau\tau}$, respectively, demonstrating that the requirements on these variables effectively distinguish signal events from the SM backgrounds.
Table \ref{tab:numdt} summarizes the event yields from background processes and reference signal points after applying the full signal region criteria. The dominant background contributions originate from $ZZ \to \tau\tau\nu\nu$, $\nu Z , Z \to \tau\tau$, $ZZ$ or $WW \to \tau\tau\nu\nu$, $ZZ \to \mu\mu\nu\nu$, $WW \to \ell\ell\nu\nu$,  $\nu Z , Z \to \mu\mu$, $\tau\tau$ and $\mu\mu$ processes.

\begin{figure*}[!htp]
  \centering
    \subfigure [systematic uncertainty = 5\%]{\includegraphics[width=.48\textwidth]{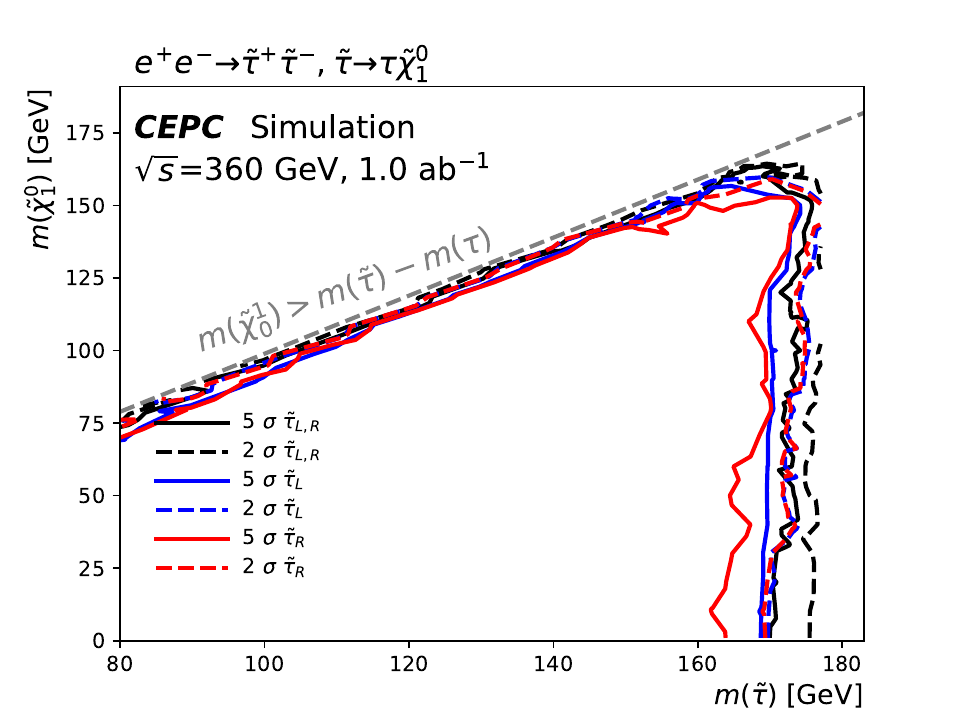}}
    \subfigure [comparison between systematic uncertainty = 0\% and 5 \%] {\includegraphics[width=.48\textwidth]{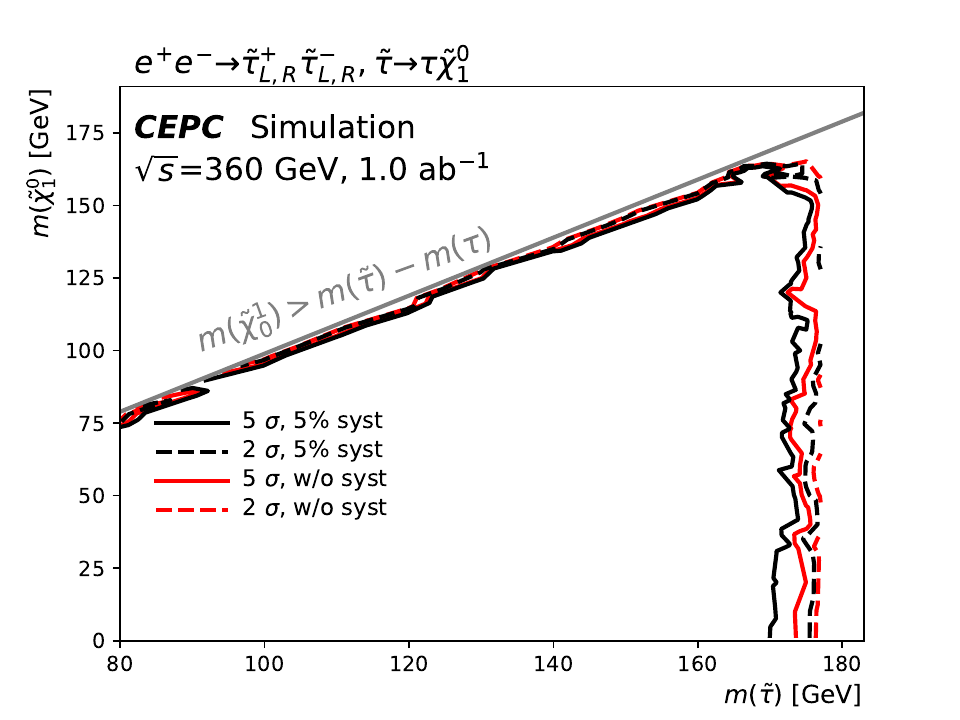}}
    \caption{The projected exclusion (2$\sigma$)  and discovery ($5\sigma$) contours for direct $\stau$ production at the CEPC.}
    \label{fig:summapdt}
\end{figure*}

Figure~\ref{fig:summapdt} presents the projected exclusion ($2\sigma$) and discovery ($5\sigma$) contours for direct $\tilde{\tau}$ production, under the assumption of 0\% systematic uncertainties (no systematics) and 5\% systematic uncertainties. The sensitivity of each signal point is derived using the optimum expected limits from SR-$\Delta M^h$, SR-$\Delta M^m$, and SR-$\Delta M^l$. To assess the effect of detector-related systematics, a conservative 5\% flat systematic uncertainty consistent with the LEP experiment is adopted. Under this framework, the discovery reach extends up to 170\,GeV for the combined left- and right-handed stau scenario, and up to 169\,GeV and 162\,GeV for purely left-handed and right-handed staus, respectively. These results indicate only limited degradation of sensitivity cause by the detector systematics.
The most sensitive signal region is selected for each individual point, and therefore, adjacent signal points in overlapping regions may be associated with different signal regions; this can result in non-smooth contours. The resulting limits extend the LEP exclusion bounds by approximately 74\,GeV and bridge the gap in the exclusion reach of LHC experiments within the compressed region, as reported in Ref.~\cite{PhysRep1116(2025)261}.

\subsection{Search for direct smuon production}
\label{subsec:searchdm}
\begin{table*}
  \centering
  \caption{Number of events in the signal regions for signal and SM backgrounds with statistical uncertainty for direct stau production.}
  \label{tab:numdt}
  \begin{tabular*}{\textwidth}{@{\extracolsep{\fill}}cccc@{}}
\hline
Process & SR-$\Delta M^h$ & SR-$\Delta M^m$ & SR-$\Delta M^l$ \\
\hline
$ZZ$ or $WW ~(\to \tau\tau\nu\nu)$             &  78$\pm$7  &    107$\pm$8&   8.2$\pm$2.3\\[0.56ex]
$\tau\tau$                                  & 15$\pm$4 & 52$\pm$7 &    0.0$\pm$0.0\\[0.56ex]
$\nu Z ~(Z \to \tau\tau)$                    & 161$\pm$10 &  165$\pm$10 & 7.5$\pm$2.2 \\[0.56ex]
$ZZ ~(\to \tau\tau\nu\nu)$                     &  239$\pm$11  &  90$\pm$7&  4.5$\pm$1.5  \\[0.56ex]
$WW ~(\to \ell\ell\nu\nu)$                           & 68$\pm$7 &  78$\pm$7 &    4.9$\pm$1.7 \\[0.56ex]
$\nu Z ~(Z \to \mu\mu)$                      & 2.5$\pm$1.3 &   15$\pm$4&   6.3$\pm$2.0\\[0.56ex]
$\mu\mu$                                    & 0.0$\pm$0.0 &5.5$\pm$2.0&  2.1$\pm$1.2\\[0.56ex]
$ZZ$ or $WW ~(\to \mu\mu\nu\nu)$               & 3$\pm$1 & 0.59$\pm$0.59&  0.0$\pm$0.0\\[0.56ex]
$ZZ ~(\to \mu\mu\nu\nu)$                       & 1.3$\pm$0.9 &0.63$\pm$0.63&  1.9$\pm$1.1\\[0.56ex]
$e\nu W~(W \to \tau\nu)$                     & 103$\pm$8  &100$\pm$8 &   5.0$\pm$1.8\\[0.56ex]
$e\nu W ~(W \to \mu\nu)$                      & 52$\pm$6  &7.0$\pm$2.2 &   0.0$\pm$0.0\\[0.56ex]
$eeZ ~(Z \to \nu\nu)$  or  $e\nu W ~(W \to e\nu)$ & 62$\pm$7 &  15$\pm$3& 0.0$\pm$0.0\\[0.56ex]
$eeZ ~(Z \to \nu\nu)$                         & 3.2$\pm$1.1 & 0.0$\pm$0.0 & 0.0$\pm$0.0\\[0.56ex]
$\nu\nu H ~(H\to$ anything)                   & 43$\pm$5 &17$\pm$3&0.7$\pm$0.7\\[0.56ex]
2-photon & 0.0$\pm$0.0 & 0.0$\pm$0.0 & 105.6$\pm$2.3 \\[0.56ex]
\hline
Total SM                            & 835$\pm$22 & 652$\pm$20&   146$\pm$5\\\hline
m($\stau$,\ninoone) = (160,50) GeV          & 946$\pm$21  &  103$\pm$7&   0.9$\pm$0.6\\[0.56ex]
m($\stau$,\ninoone) = (160,110) GeV          &  958$\pm$21  &  909$\pm$20&    18$\pm$3 \\[0.56ex]
m($\stau$,\ninoone) = (160,150) GeV         & 0.0$\pm$0.0&3.1$\pm$1.2 &   151$\pm$8\\
\hline      
\end{tabular*}
\end{table*}

\begin{table*}
  \centering
  \caption{Summary of selection requirements for direct smuon production signal regions. $\Delta M$ denotes the mass difference between the $\smu$ and LSP.}
  \label{tab:SRdm}
  \begin{tabular*}{\textwidth}{@{\extracolsep{\fill}}ccc@{}}
\hline
SR-$\Delta M^h$ & SR-$\Delta M^m$ & SR-$\Delta M^l$ \\
\hline
%\multicolumn{3}{c}{ Preselection: Events have two OS taus whose $E_\tau$ \textgreater 0.5 GeV} \\
$E_{\mu1,2}>60$ GeV & $E_{\mu1,2}<80$ GeV  &$E_{\mu1,2}>2.5$ GeV \\[0.56ex]
$ E_{\mu1,2}\in (60-70, >70)$ GeV& $E_{\mu1,2}\in (<35,35-80)$ GeV &$\Delta \Phi(\mu,recoil)<$2.6\\[0.56ex]
%$E_{\mu1,2}(>60 GeV) \in (60-70, 70-)$ GeV& $E_{\mu1,2}(<80 GeV) \in (-35,35-80)$ GeV &-\\[0.56ex]
$\Delta R(\mu,recoil) < $ 2.8 &$1.9<\Delta R(\mu,recoil)<$2.9 & $1.9<\Delta R(\mu,recoil)<$2.8\\[0.56ex]
$M_{\mu\mu}<$87 GeV $||$  $95<M_{\mu\mu}<130$ GeV & $M_{\mu\mu}<80$ GeV & $M_{\mu\mu}>5$ GeV \\[0.56ex]
$M_{recoil}>100$ GeV & -- & $M_{recoil}>340$ GeV\\[0.56ex]
$M_{extra}<$15 GeV & $M_{extra}<$10 GeV& $M_{extra}<$10 GeV\\[0.56ex]
\hline
\end{tabular*}
\end{table*}

% \resizebox{\textwidth}{!}{
\begin{table*}
    \centering
    \caption{Number of events in the signal regions for signal and SM backgrounds with statistical uncertainty for direct smuon production.}
    \label{tab:numdm}
\resizebox{\textwidth}{!}{ 
\begin{tabular*}{1.1\textwidth}{c|c|c|c|c|c|c|c|c|c}
\cline{1-10}
% \hline
\multirow{3}{*}{Process}  & \multicolumn{4}{|c|}{SR-$DM^h$}        & \multicolumn{4}{|c|}{SR-$DM^m$}      & \multicolumn{1}{|c}{\multirow{4}{*}{SR-$DM^l$}}    \\[0.6ex]
\cline{2-9}
 & \multicolumn{4}{|c|}{$E_{\mu 1,2}$ \textgreater 60 GeV} & \multicolumn{4}{|c|}{$E_{\mu 1,2}$   \textless 80 GeV} &  \\[0.6ex]
\cline{2-9}
% & \begin{tabular}[c]{@{}c@{}}60\textless $E_{\mu 1}$\textless70\\    \\ 60\textless $E_{\mu 2}$\textless70\end{tabular} & \begin{tabular}[c]{@{}c@{}}60\textless $E_{\mu 1}$\textless70\\    \\ $E_{\mu 2}$\textgreater{}70\end{tabular} & \begin{tabular}[c]{@{}c@{}}$E_{\mu 1}$\textgreater{}70\\    \\ 60\textless $E_{\mu 2}$\textless70\end{tabular} & \begin{tabular}[c]{@{}c@{}}$E_{\mu 1}$\textgreater{}70\\    \\ $E_{\mu 2}$\textgreater{}70\end{tabular} & \begin{tabular}[c]{@{}c@{}}$E_{\mu 1}$\textless{}35\\    \\ $E_{\mu 2}$\textless{}35\end{tabular} & \begin{tabular}[c]{@{}c@{}}$E_{\mu 1}$\textless{}35\\    \\ 35\textless$E_{\mu 2}$\textless 80\end{tabular} & \begin{tabular}[c]{@{}c@{}}35\textless$E_{\mu 1}$\textless 80\\    \\ $E_{\mu 2}$\textless{}35\end{tabular} & \begin{tabular}[c]{@{}c@{}}35\textless$E_{\mu 1}$\textless 80\\    \\ 35\textless$E_{\mu 2}$\textless 80\end{tabular} &  \\[0.6ex]
 & \begin{tabular}[c]{@{}c@{}c@{}} Bin1 \\  60\textless $E_{\mu 1}$\textless70\\    \\ 60\textless $E_{\mu 2}$\textless70\end{tabular} & \begin{tabular}[c]{@{}c@{}c@{}} Bin2 \\60\textless $E_{\mu 1}$\textless70\\    \\ $E_{\mu 2}$\textgreater{}70\end{tabular} & \begin{tabular}[c]{@{}c@{}c@{}} Bin3 \\$E_{\mu 1}$\textgreater{}70\\    \\ 60\textless $E_{\mu 2}$\textless70\end{tabular} & \begin{tabular}[c]{@{}c@{}c@{}} Bin4 \\$E_{\mu 1}$\textgreater{}70\\    \\ $E_{\mu 2}$\textgreater{}70\end{tabular} & \begin{tabular}[c]{@{}c@{}c@{}} Bin1 \\$E_{\mu 1}$\textless{}35\\    \\ $E_{\mu 2}$\textless{}35\end{tabular} & \begin{tabular}[c]{@{}c@{}c@{}} Bin2 \\$E_{\mu 1}$\textless{}35\\    \\ 35\textless$E_{\mu 2}$\textless 80\end{tabular} & \begin{tabular}[c]{@{}c@{}c@{}} Bin3 \\35\textless$E_{\mu 1}$\textless 80\\    \\ $E_{\mu 2}$\textless{}35\end{tabular} & \begin{tabular}[c]{@{}c@{}c@{}} Bin4 \\35\textless$E_{\mu 1}$\textless 80\\    \\ 35\textless$E_{\mu 2}$\textless 80\end{tabular} &  \\[0.6ex]

\cline{1-10}
%\smallskip 
ZZ or WW ~($\to$ $\tau\tau\nu\nu)$ &4.4$\pm$1.7 &2.5$\pm$1.3 &0.6$\pm$ 0.6&0.0$\pm$0.0&65$\pm6$ &10.0$\pm$3.0&16.0$\pm$3.0 &6.9$\pm2.1$ & 2.5$\pm$1.3  \\[0.6ex]
$\tau\tau$     & $9.0\pm$2.5 & 0.0$\pm$0.0 & 2.1$\pm$1.2 & 0.0$\pm$0.0 & 132$\pm$10 & 21$\pm$4 & 33$\pm$5 & 23$\pm$4 & 2.8$\pm$1.4 \\[0.6ex]
$\nu$Z~(Z $\to$ $\tau\tau)$ &0.0$\pm$0.0 &0.0$\pm$0.0 &0.0$\pm$0.0 &0.0$\pm$0.0&54$\pm$6 &3.8$\pm$1.5 &5.0$\pm$1.8 &3.8$\pm$1.5 & 3.8$\pm$1.5      \\[0.6ex]
ZZ ~($\to$ $\tau\tau\nu\nu)$  &0.0$\pm$0.0 &0.5$\pm$0.5 &0.5$\pm$0.5 &0.0$\pm$0.0 &27$\pm$4 &4.0$\pm1.4$ &5.5$\pm$1.7 &5.0$\pm$1.6 & 3.0$\pm$1.2 \\[0.6ex]
WW ~($\to$ $\ell\ell\nu\nu)$ &39$\pm$5 &12.0$\pm$3.0 &11.0$\pm$ 3.0& 2.5$\pm$1.2& 228$\pm$12 &58$\pm$6 & 
97$\pm$8 &61$\pm$6 & 11.0,2.6 \\[0.6ex]
$\nu$Z~(Z $\to$ $\mu\mu)$  &3.8$\pm$1.5 &1.9$\pm$1.1 &5.6$\pm$1.9 &8.1$\pm$2.3& 132$\pm$9 &18.0$\pm$3.0 &39$\pm$5 &16.0$\pm$3.0 &15.6$\pm$3.1      \\[0.6ex]
$\mu\mu$  &16.0$\pm$3.0 &12.0$\pm$3.0 &15.0$\pm$3.0 &21$\pm$4&337$\pm$15 &31$\pm$5 &40$\pm$5 &33$\pm$5 & 36$\pm$5        \\[0.6ex]
ZZ or WW ~($\to$ $\mu\mu\nu\nu)$  &112$\pm$8 &73$\pm$7 &81$\pm$7 &46$\pm$5& 329$\pm$14 &134$\pm$9 &271$\pm$13 &135$\pm$ 9& 11.7$\pm$2.6      \\[0.6ex]
ZZ ~($\to$ $\mu\mu\nu\nu)$  &1.9$\pm$1.1 &8.1$\pm$2.2 &9.4$\pm$2.4 & 24$\pm$4& 64$\pm$6 &8.1$\pm$2.3 &23$\pm$4 &8.1$\pm$2.4 & 4.4$\pm$1.7      \\[0.6ex]
$\nu\nu$H~(H $\to$ anything)   &0.0$\pm$0.0 &0.8$\pm$0.8 & 0.0$\pm$0.0 & 0.8$\pm$0.8&12.0$\pm$3.0 &5.0$\pm$2.0 & 5.0$\pm$2.0&3.3$\pm$1.7 &0.0$\pm$0.0 \\[0.6ex]
$t\bar{t}$  &0.07$\pm$ 0.07&0.0$\pm$0.0 &0.0$\pm$0.0 &0.0$\pm$0.0&0.07$\pm$ 0.07&0.0$\pm$0.0 &0.07$\pm$0.07 &0.0$\pm$ 0.0 &0.0$\pm$ 0.0  \\[0.6ex]
2-photon & 0.0$\pm$ 0.0 & 0.0$\pm$ 0.0 & 0.0$\pm$ 0.0 & 0.0$\pm$ 0.0 & 0.0$\pm$ 0.0 & 0.0$\pm$ 0.0 &0.0$\pm$ 0.0&0.0$\pm$ 0.0& 106.1 $\pm$ 2.3 \\[0.6ex]

\cline{1-10}
Total SM  &185$\pm$ 11&111$\pm$8 &125$\pm$9 &103$\pm$8&1379$\pm$ 29&293$\pm$13 &535$\pm$18 &294$\pm$13 & 186 $\pm$ 7                 \\[0.6ex]
\cline{1-10}
m($\smu$,\ninoone) = (170,  30) GeV  &163$\pm$ 5&200$\pm$6&219$\pm$6 &192$\pm$6& 8.6$\pm$1.2 &15.0$\pm$2.0 &57.0$\pm$3.0 & 0.66$\pm$0.33 &4.3$\pm$ 0.8            \\[0.6ex]
m($\smu$,\ninoone) = (170, 100) GeV  &453$\pm$9 &311$\pm$7 &300$\pm$7 &46.0$\pm$3.0& 3.0$\pm$0.7 & 10$\pm$1 & 1870$\pm$18 &43.0$\pm$3.0 &1.2$\pm$0.4              \\[0.6ex]
m($\smu$,\ninoone) = (170, 165) GeV    &0.0$\pm$0.0 &0.0$\pm$0.0 &0.0$\pm$0.0 &0.0$\pm$0.0& 2310$\pm$20 &0.0$\pm$0.0 &0.0$\pm$0.0 &0.0$\pm$ 0.0&1158$\pm$  14   \\[0.6ex]
\cline{1-10}
\end{tabular*}
}
\end{table*}

% }

\begin{figure*}[!htp]
  \centering
    \subfigure [$M_{recoil}$] {\includegraphics[width=.48\textwidth]{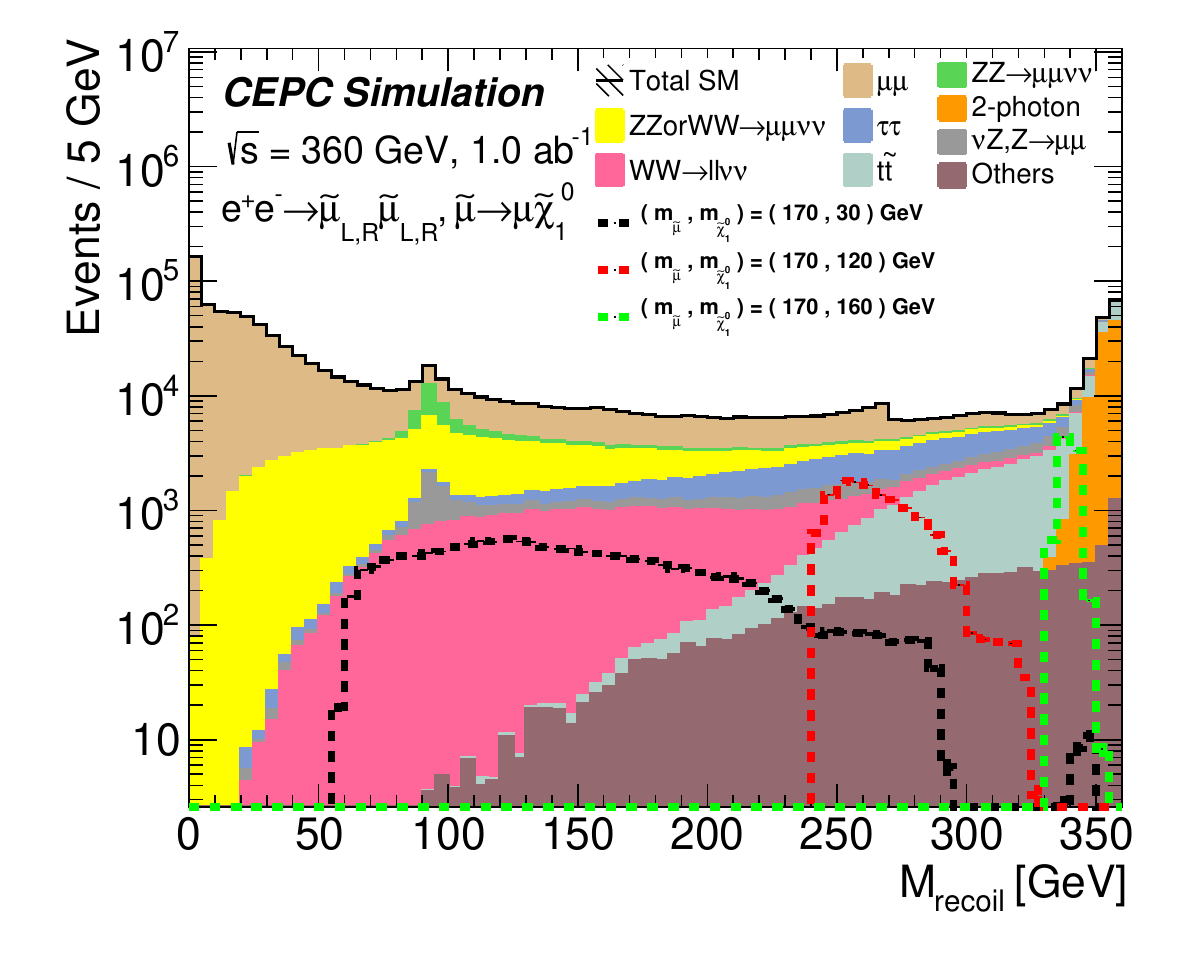}}
    \subfigure [$M_{\mu\mu}$] {\includegraphics[width=.48\textwidth]{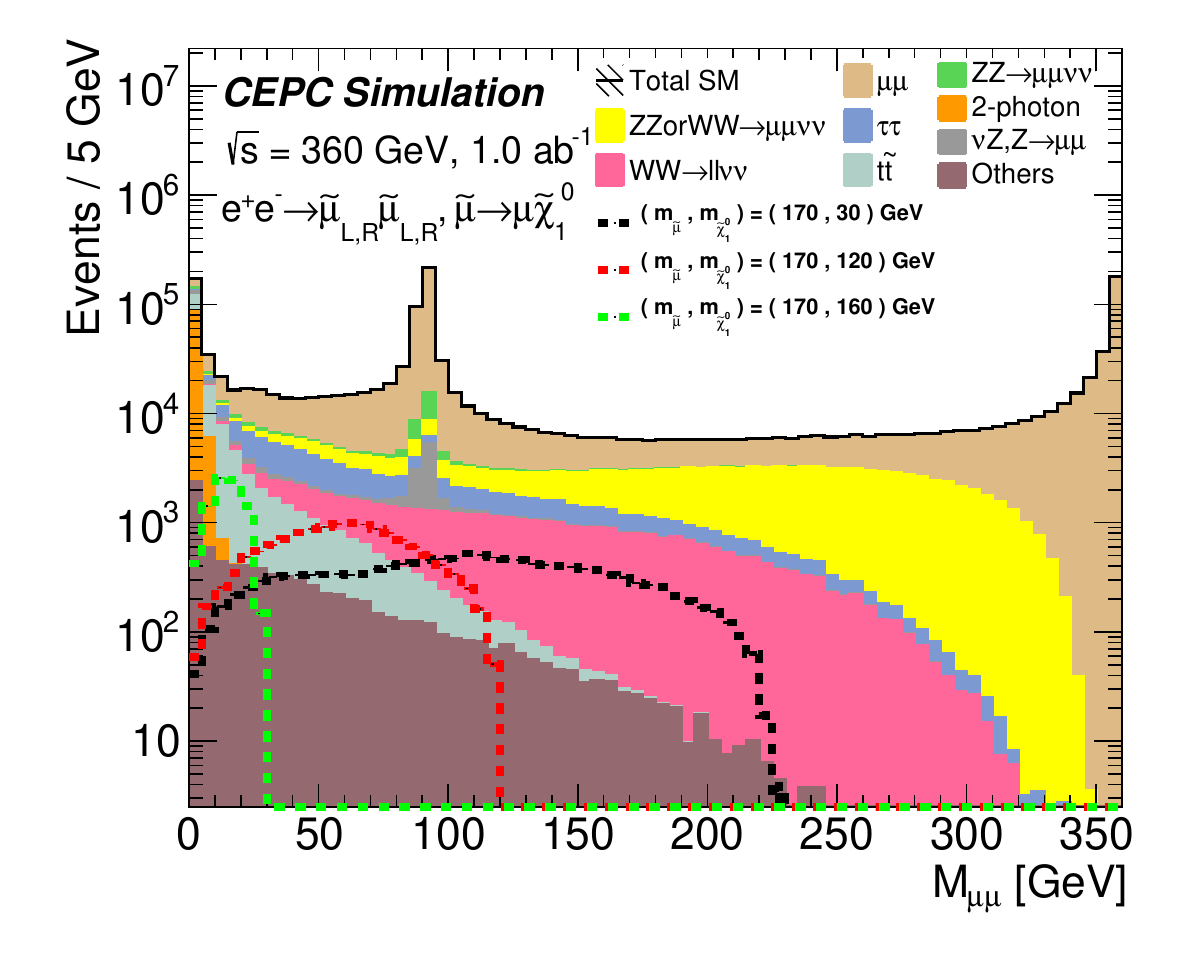}}
    \subfigure [$E_{\mu1}$] {\includegraphics[width=.48\textwidth]{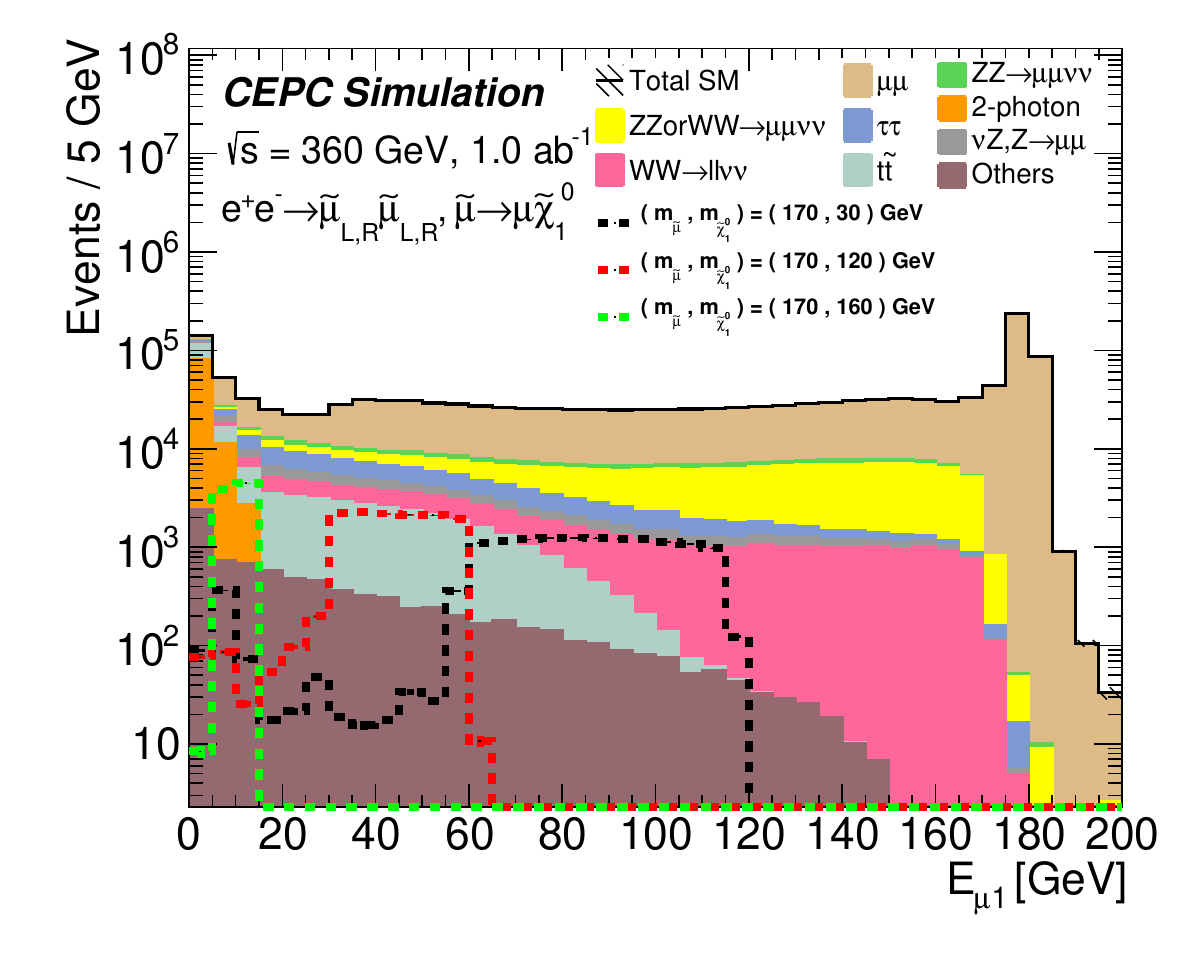}}
    \subfigure [$M_{extra}$] {\includegraphics[width=.48\textwidth]{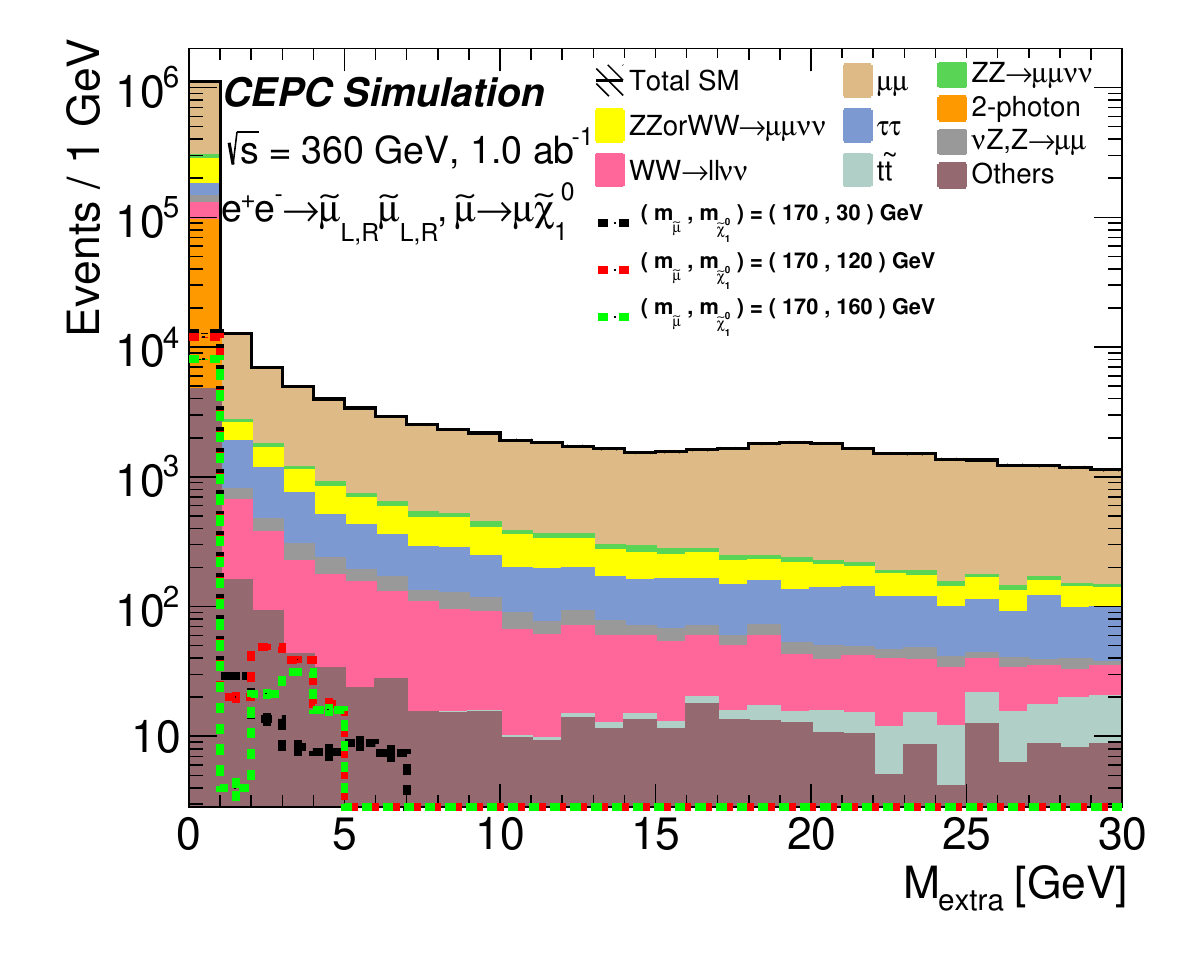}}
    \caption{Kinematic distributions for direct smuon pair production after applying the preselection. The stacked histograms show the expected SM background. The distributions of three SUSY reference points are shown as dashed lines.}
    \label{fig:nm1dm}
  \end{figure*}

For smuon pair production, events that contain exactly two OS muons are selected; each muon satisfies an energy threshold exceeding 2.5 GeV and an absolute pseudorapidity less than 2.5. This criterion is implemented to suppress the two-photon background.
The kinematic distributions of the variables used in the analysis, after applying the above selection criteria, are presented in Figure~\ref{fig:nm1dm}, which demonstrates strong discrimination power between signal events and the SM background processes.

\begin{figure*}[!htp]
  \centering
    \subfigure [$E_{\mu1}$] {\includegraphics[width=.48\textwidth]{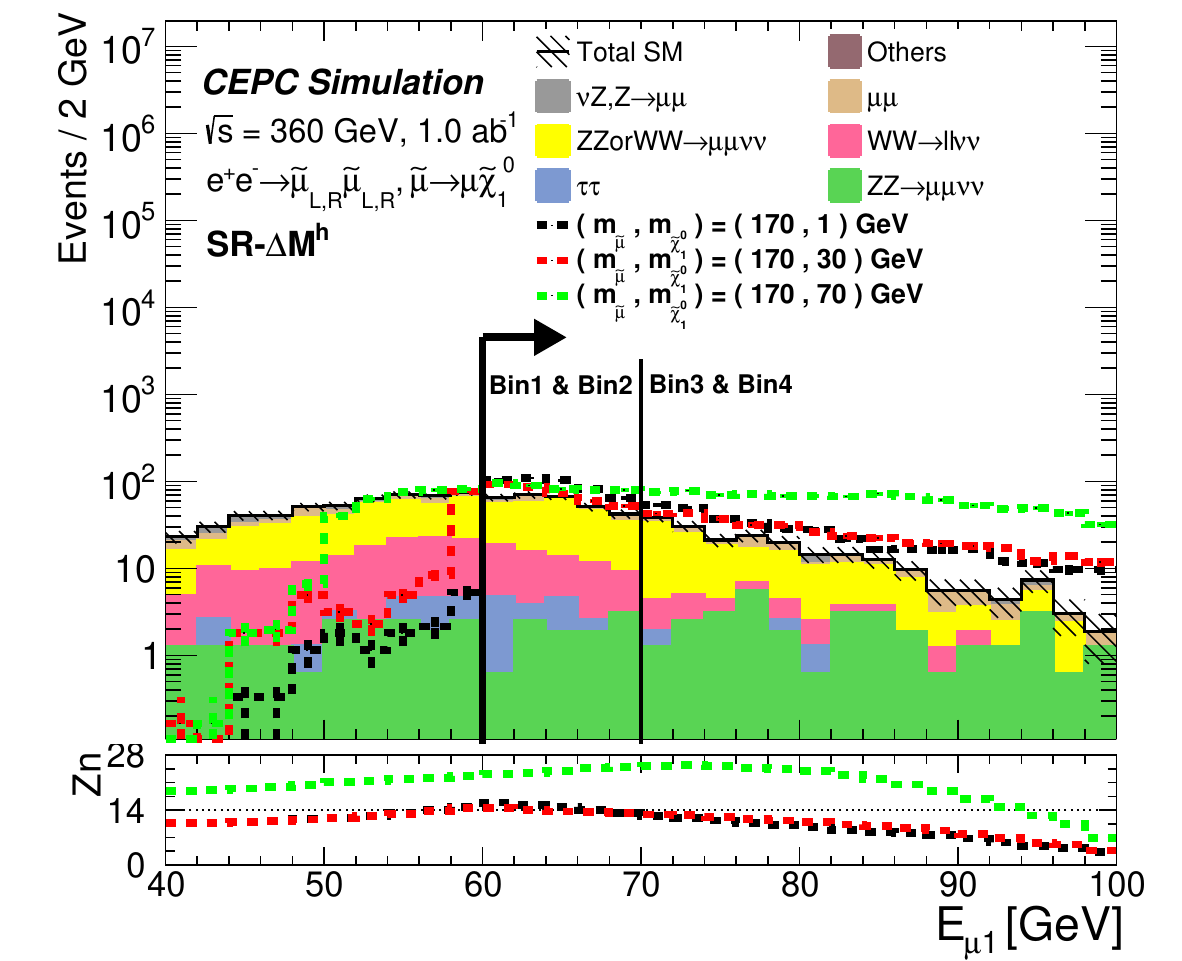}}
    \subfigure [$M_{\mu\mu}$] {\includegraphics[width=.48\textwidth]{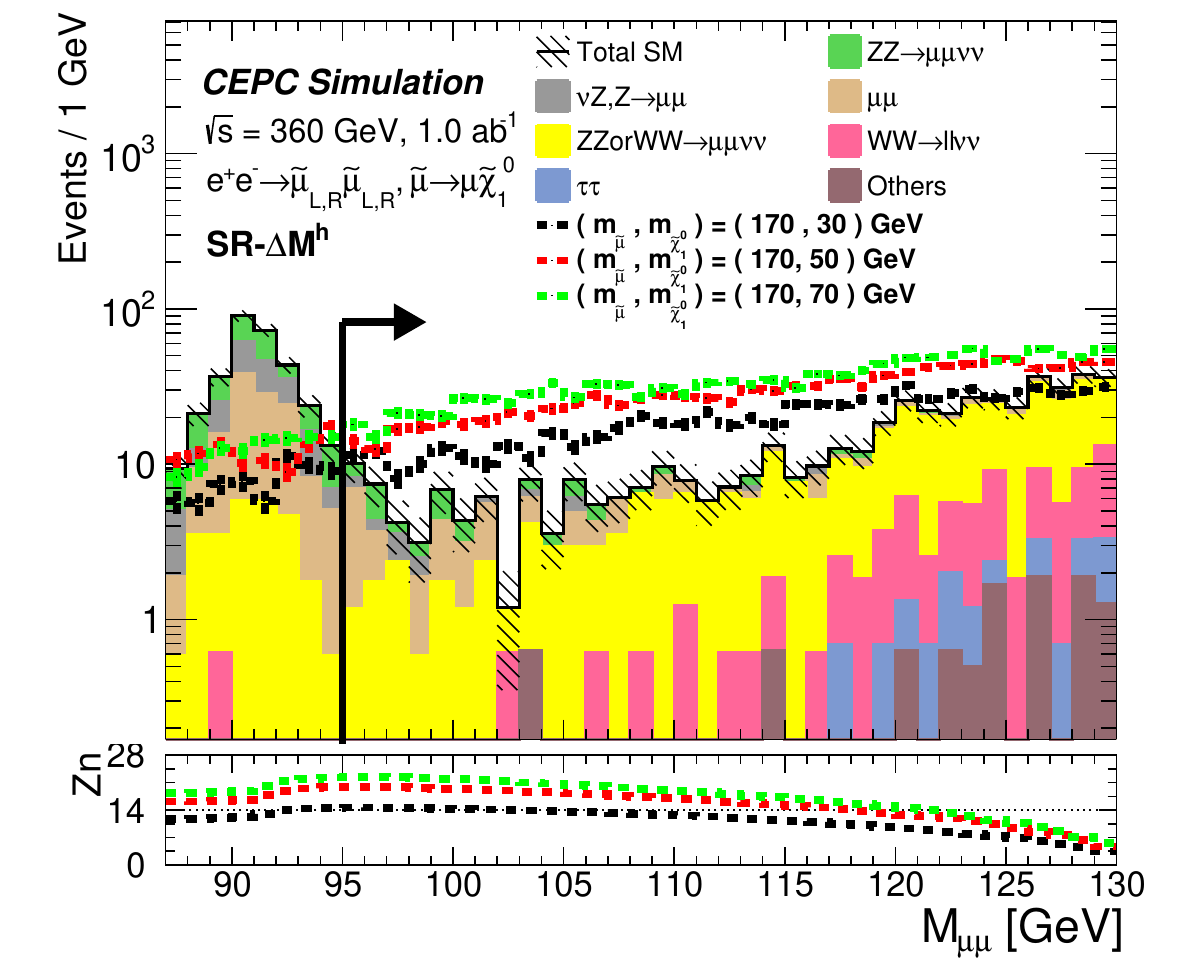}}
    \subfigure [$E_{\mu1}$] {\includegraphics[width=.48\textwidth]{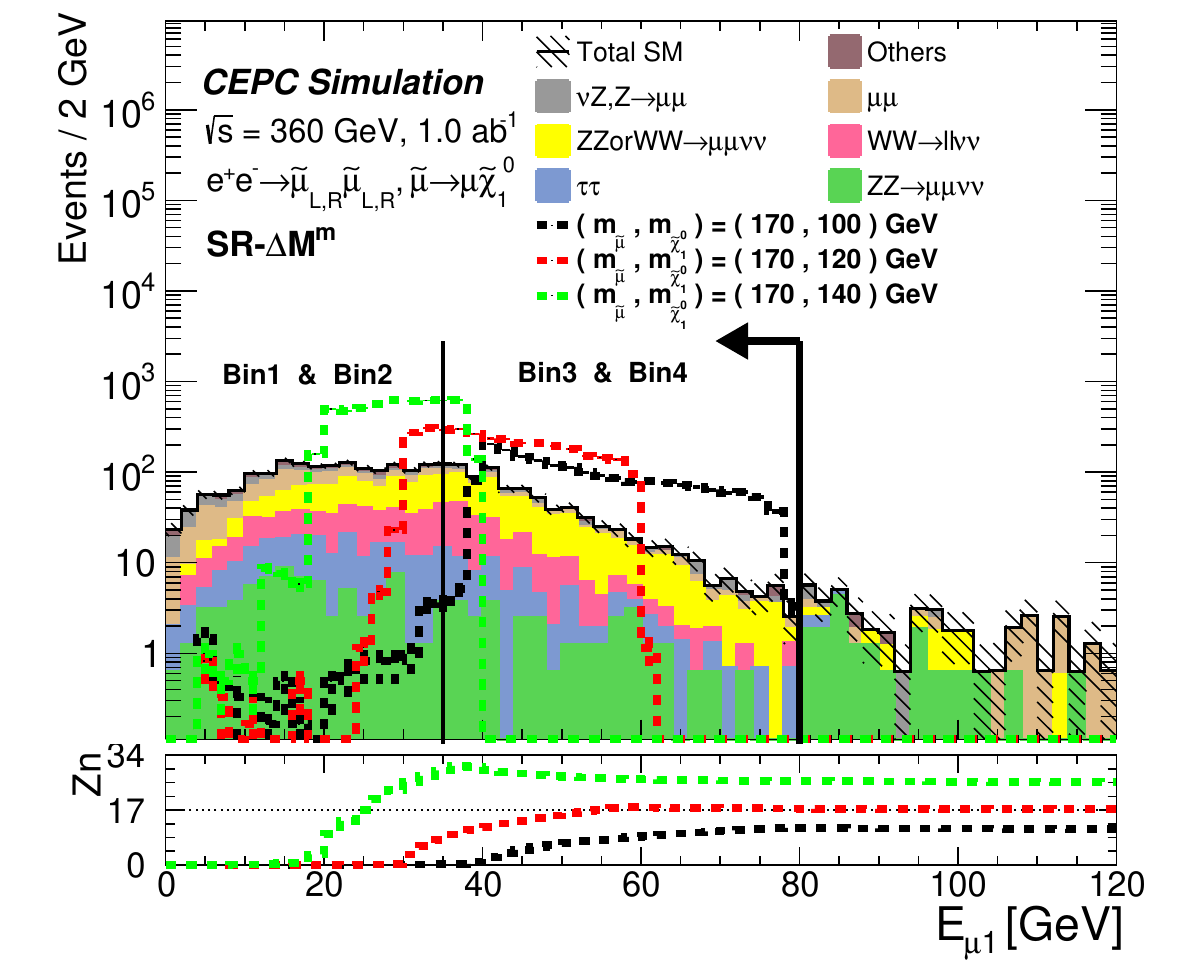}}
    \subfigure [$M_{\mu\mu}$] {\includegraphics[width=.48\textwidth]{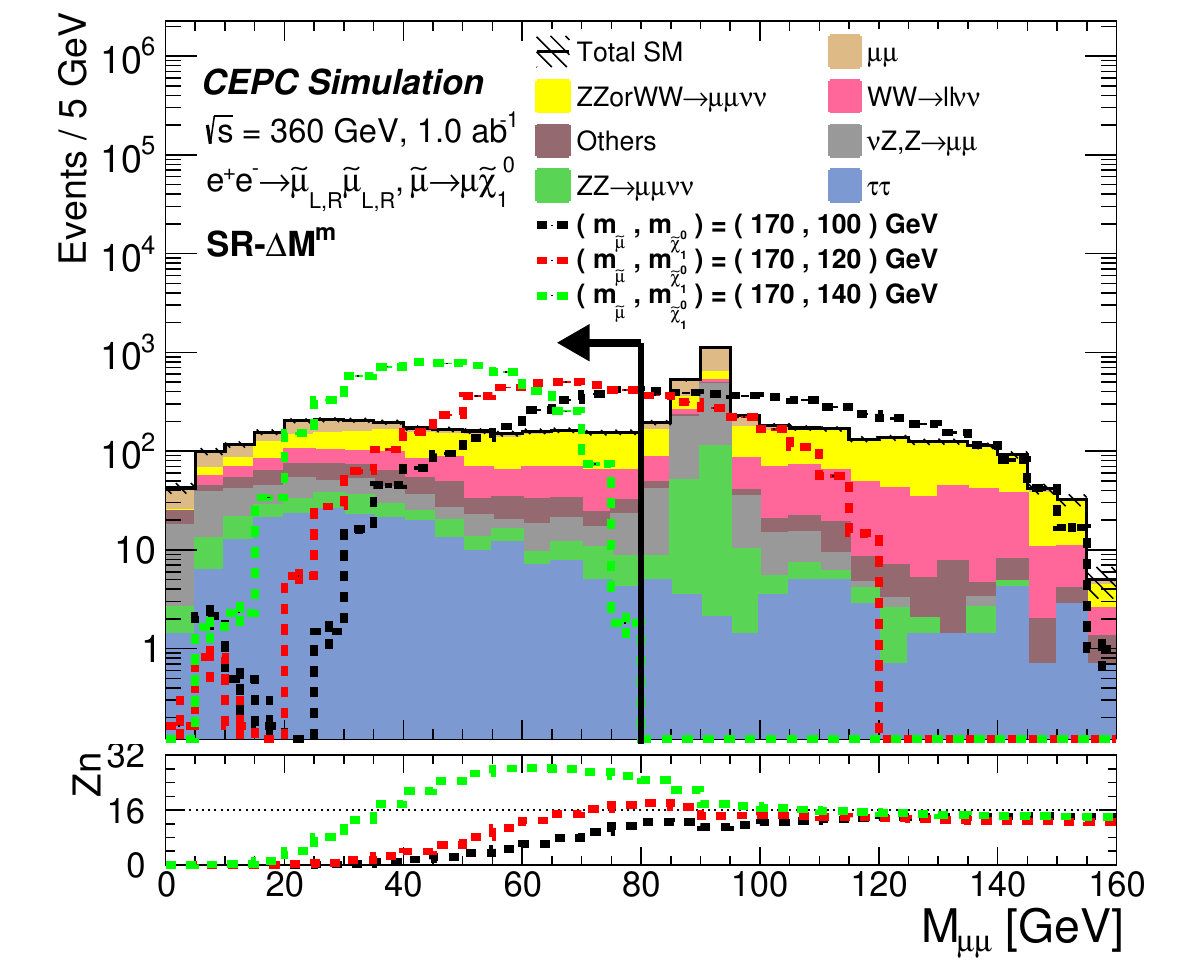}}
    \subfigure [$M_{recoil}$] {\includegraphics[width=.48\textwidth]{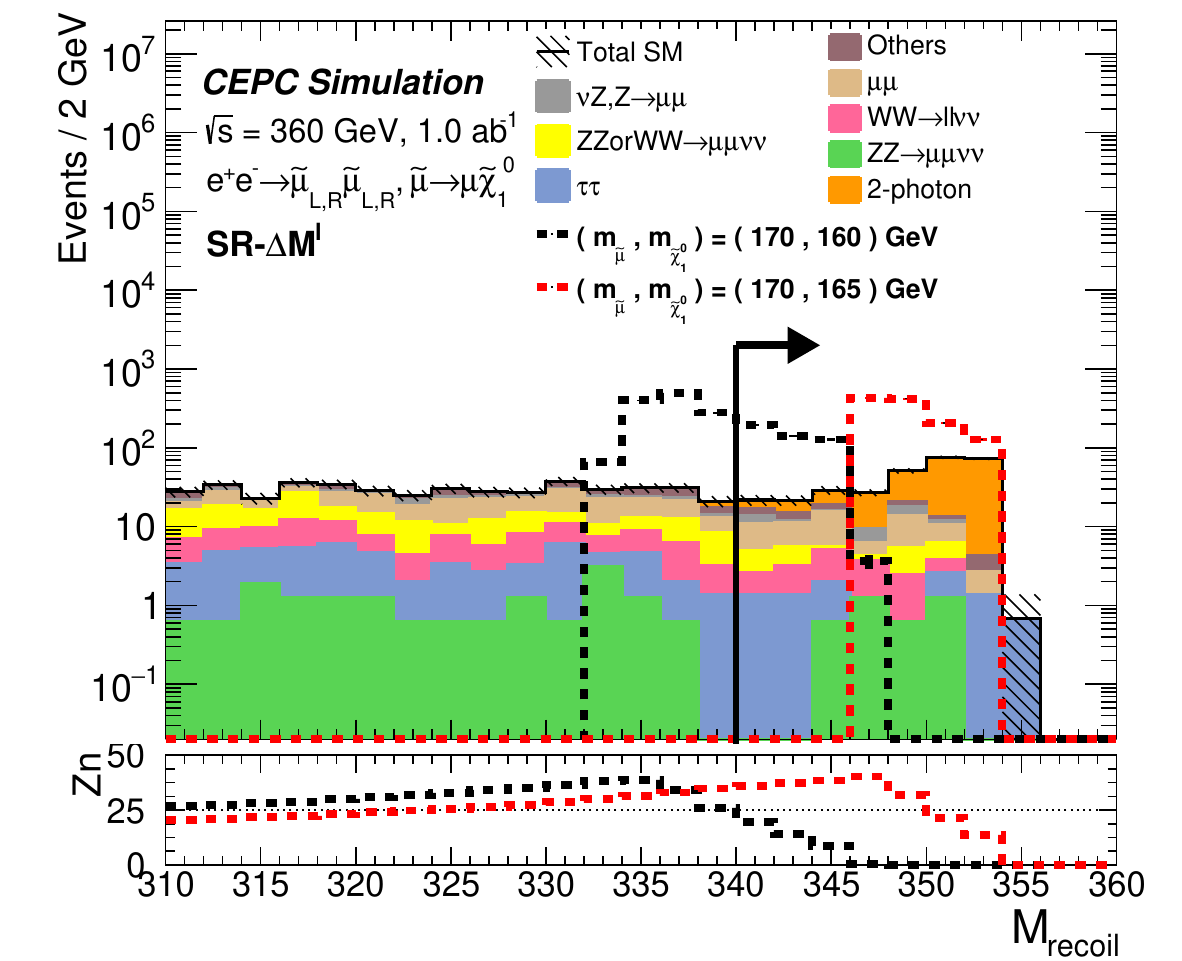}}
    \subfigure [$M_{\mu\mu}$] {\includegraphics[width=.48\textwidth]{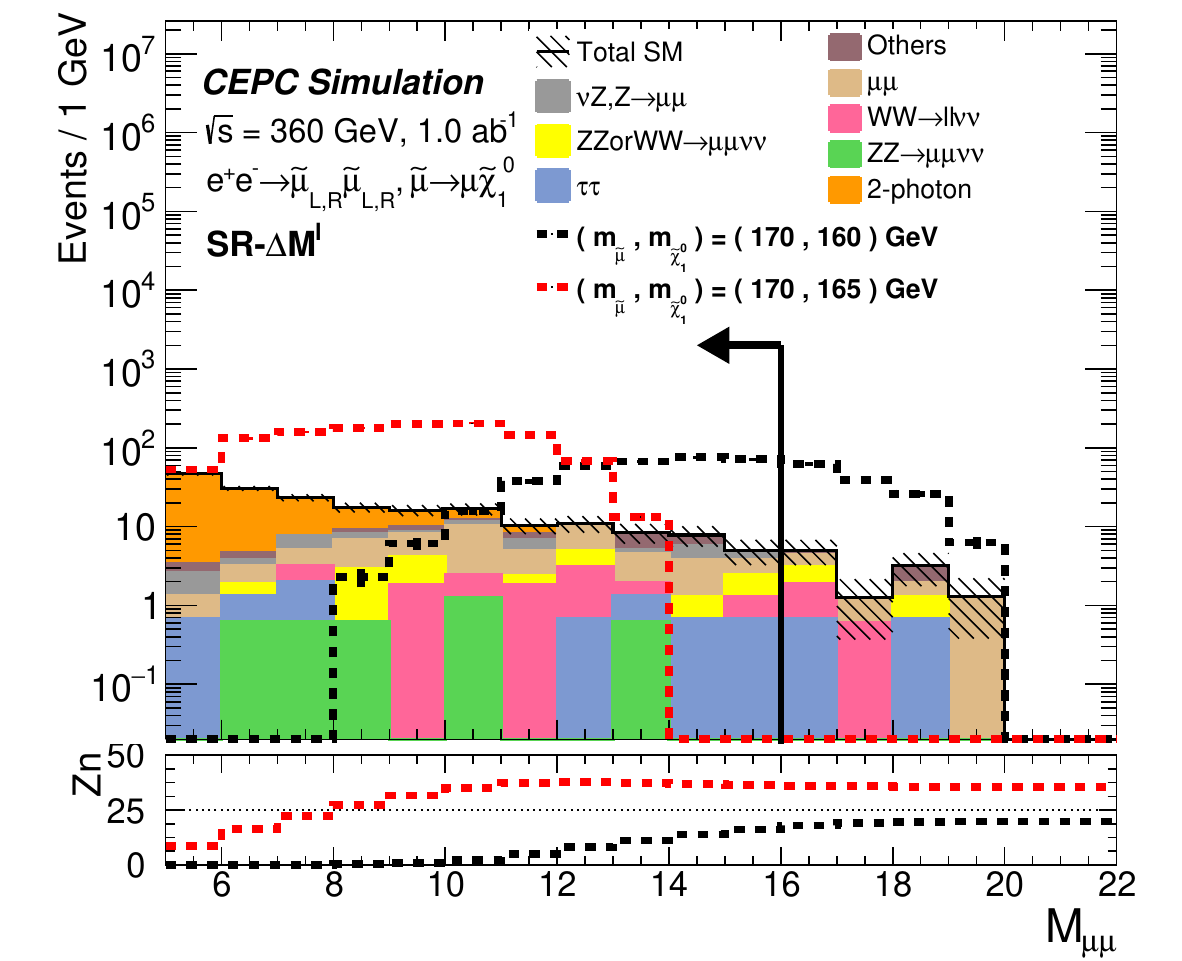}}
    \caption{"N-1" distributions after signal region requirements for direct smuon pair production. All signal region requirements are applied except for the variable shown. The stacked histograms show the expected SM backgrounds. The distributions from SUSY reference points are shown as dashed lines. The lower pad is the sensitivity Zn calculated with a statistical uncertainty and a 5\% flat systematic uncertainty.}
    \label{fig:nm0dm}
  \end{figure*}

Similar to stau pair production, three SRs are designed to cover different mass splittings between \smu and \ninoone ($\Delta M$). SR-$\Delta M^h$ encompasses the high-$\Delta M$ region, SR-$\Delta M^m$ covers the medium-$\Delta M$ region, and SR-$\Delta M^l$ spans the low-$\Delta M$ region. 
To enhance signal sensitivity, the SR-$\Delta M^h$ and SR-$\Delta M^m$ regions are further subdivided into intervals based on the energies of the leading and sub-leading muons (E$_{\mu1}$ and E$_{\mu2}$, collectively denoted as E$_{\mu1,2}$), where $\mu1$ ($\mu2$) refers to the muon with the highest (second-highest) energy.
The definitions of signal regions are summarized in Table \ref{tab:SRdm}. 
Selections on E$_{\mu1,2}$ are applied to reject the  $\tau\tau$ and $\nu$Z background processes. Cuts on $\Delta R(\mu, recoil)$ are used to suppress $\tau\tau$, $\mu\mu$ and ZZ backgrounds, while cuts on the dimuon invariant mass $M_{\mu\mu}$ target the suppression of WW, $\mu\mu$, two-photon and other Z-related backgrounds.
Consistent with signal topology, where most signal events exhibit a large recoil mass, a lower cut on the invariant mass of the recoil system ($M_{recoil}$) is implemented to reject $\mu\mu$, Z or W mixing processes, and other SM processes lacking significant recoil mass. In addition, the invariant mass of all extra reconstructed visible particles (with energies above 0.5 GeV, excluding two oppositely charged muon tracks), denoted as $M_{extra}$, is utilized to strongly suppress the $t\bar{t}$ background.

The kinematic distributions of $M_{recoil}$, $M_{\mu\mu}$, $E_{\mu1}$ and $\Delta R$ $(\mu,recoil)$ after applying all signal region requirements except those on the variable being displayed are presented in Figure \ref{fig:nm0dm}. The lower panel of each plot shows the expected sensitivity Zn as a function of $M_{recoil}$, $M_{\mu\mu}$, $E_{\mu1}$ and $\Delta R$ $(\mu,recoil)$, respectively,which demonstrates that selections on $M_{recoil}$ and $M_{\mu\mu}$ effectively distinguish SUSY signal events from the SM background processes.
Table \ref{tab:numdm} summarizes the event yields from all background processes and reference signal points after applying the full signal region criteria. The dominant background contributions originate from $ZZ$ or $WW \to \mu\mu\nu\nu$, $\mu\mu$, $WW\to\ell\ell\nu\nu$, $ZZ\to\mu\mu\nu\nu$, $\tau\tau$, $ZZ$ or $WW \to \tau\tau\nu\nu$ and $\nu Z~(Z\to\tau\tau$) processes. 

\begin{figure*}[!htp]
  \centering
  {\includegraphics[width=.8\textwidth]{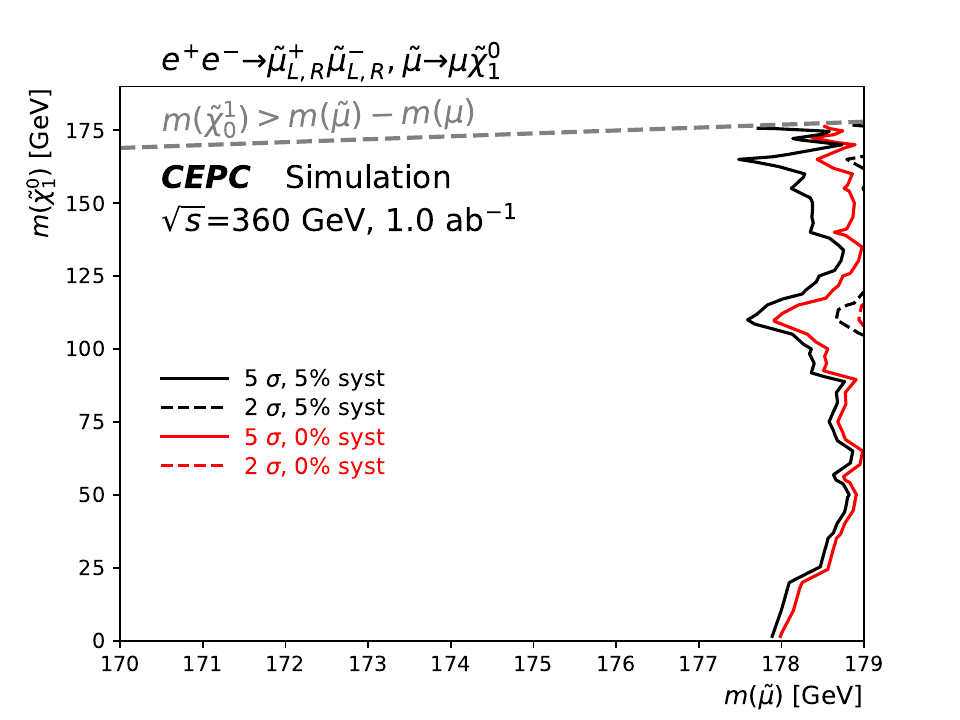}}
    \caption{The projected exclusion ($2\sigma$) and discovery ($5\sigma$) contours for direct \smu production at the CEPC, under the assumptions of 0\% and 5\% systematic uncertainties.}
    \label{fig:summapdm}
\end{figure*}

Figure~\ref{fig:summapdm} shows the projected exclusion ($2\sigma$) and discovery ($5\sigma$) contours for direct smuon production, assuming 0\% and 5\% systematic uncertainties. The sensitivity at each signal point is derived using the optimal expected limits from SR-$\Delta M^h$, SR-$\Delta M^m$, and SR-$\Delta M^l$. Under the assumption of a 5\% flat systematic uncertainty, the discovery reach extends up to 178\,GeV in $\tilde{\mu}$ mass, with minimal degradation attributed to detector-related systematics. The most sensitive signal region is selected for each individual point, and therefore, adjacent signal points in overlapping regions may be associated with different signal regions, potentially resulting in non-smooth contours. This result extends the LEP exclusion limits by approximately 79\,GeV and bridges the gap in the compressed mass region left by LHC experiments, as documented in Ref.~\cite{PhysRep1116(2025)261}.

\begin{figure*}[!htb]
\centering
\includegraphics[width=.8\textwidth]{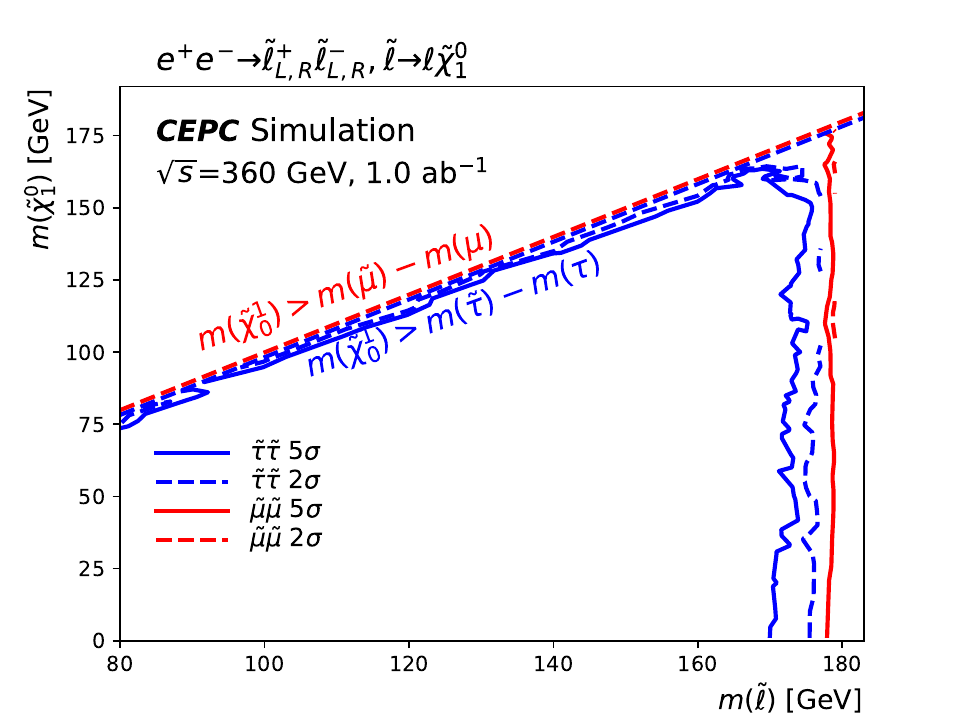}
\caption{The projected exclusion ($2\sigma$) and discovery ($5\sigma$) contours for direct \stau and direct \smu production at the CEPC, under the assumptions of 5\% systematic uncertainties.}
  \label{fig:sum}
\end{figure*}

\subsection{Summary of slepton search}
\label{subsec:sumsl}

Figure~\ref{fig:sum} presents the projected exclusion and discovery contours at the CEPC for direct $\tilde{\tau}$ and $\tilde{\mu}$ production, assuming a 5\% flat systematic uncertainty. For direct stau production, the discovery reach extends up to 170\,GeV in stau mass for the combined left- and right-handed scenario, with respective reaches of 169 and 162\,GeV for purely left-handed and right-handed staus. For direct smuon production, the discovery potential reaches up to 178\,GeV in smuon mass. Given the similar design and performance characteristics of future $e^+e^-$ collider facilities, these results serve as a valuable reference for other proposed experiments with comparable center-of-mass energies and target luminosities such as the ILC~\cite{Behnke:2013lya} and FCC-ee~\cite{Gomez-Ceballos:2013zzn}.

\section{Conclusion}
\label{sec:conclusion}
Searches for direct slepton pair production are performed at the CEPC using fully simulated MC samples, assuming a center-of-mass energy of $\sqrt{s} = 360$\,GeV and an integrated luminosity of 1.0\,ab$^{-1}$. The mass reach for $\tilde{\tau}$ ($\tilde{\mu}$) extends approximately 74 (79)\,GeV beyond the previous limits set by the LEP in the high-mass region. Furthermore, the analysis exhibits sensitivity to the compressed mass region, where the mass difference between $\tilde{\tau}$ ($\tilde{\mu}$) and the LSP is small. This regime remains challenging for ATLAS and CMS to probe effectively. This MC study provides compelling motivation for upgrading the center-of-mass energy of the CEPC from 240\,GeV to 360\,GeV, enhancing its capability to search for sleptons.

\section{Acknowledgments}
\label{sec:acknowledgments}
The authors are grateful to Manqi Ruan, Chengdong Fu, Gang Li, Xianghu Zhao, Ronggang Ping and Dan Yu for their valuable assistance with CEPC simulations and to Lorenzo Feligioni for helpful input on polishing and improving the manuscript.
This study was supported by the State Key Program of National Natural Science of China (Grant No. 2018YFA 0404000 and Grant No. 12135013).

\bibliographystyle{spphys}
\bibliography{DSL.bib}
\end{document}